\documentclass[12pt]{article}

\usepackage{natbib}
\usepackage{graphicx}
\usepackage{epstopdf}
\usepackage{secdot}
\usepackage{color}
 \bibpunct[, ]{(}{)}{,}{a}{}{,}%
\usepackage{xcolor}
\usepackage{xkeyval}
\usepackage{tikz}
\usepackage{amsmath,amssymb,amsthm}

\newtheorem{observation}{Observation}
\newtheorem{definition}{Definition}
\newtheorem{lemma}{Lemma}

\newcommand{\ie}{\emph{i.e.,   }}
\newcommand{\eg}{\emph{e.g.,   }}

\newenvironment{hangref}
  {\begin{list}{}{\setlength{\itemsep}{4pt}
  \setlength{\parsep}{0pt}\setlength{\leftmargin}{+\parindent}
  \setlength{\itemindent}{-\parindent}}}{\end{list}}

\newtheorem{theorem}{Theorem}

\begin{document}

\begin{center}

{\LARGE Capacity Allocation and Pricing Strategies for Wireless
Femtocell Services}\\[12pt]


\footnotesize


\mbox{\large Lingjie Duan}\\
Department of Information Engineering,
The Chinese University of Hong Kong
{\mbox{dlj008@ie.cuhk.edu.hk}}\\[6pt]

\mbox{\large Biying Shou}\\
Department of Management Sciences, City
University of Hong Kong, Hong Kong,
\mbox{biying.shou@cityu.edu.hk}\\[6pt]

\mbox{\large Jianwei Huang}\\
Department of Information Engineering, The Chinese University of Hong Kong,
\mbox{jwhuang@ie.cuhk.edu.hk}

\normalsize

\end{center}

\baselineskip 20pt plus .3pt minus .1pt


\noindent Indoor cell phone users often suffer from poor connectivity. One promising solution, femtocell technology, has been rapidly developed and
deployed over the past few years. One of the biggest challenges for femtocell deployment is lack of a clear business model.
This paper investigates the economic incentive for the
cellular operator (also called macrocell operator) to enable
femtocell service by leasing spectrum resource to an independent
femtocell operator. On the one hand, femtocell services can increase
communication service quality and thus increase the efficiency of the spectrum
resource. On the other hand, femtocell services may introduce more competition to the market. We model the interactions between a macrocell
operator, a femtocell operator, and users as a three-stage dynamic
game, and derive the equilibrium pricing and capacity allocation decisions. We show that when spectrum resources are very limited, the macrocell operator has incentive to lease spectrum to femtocell
operators, as femtocell service can provide access to more users and efficiently increase the coverage. However, when the total spectrum resource is large, femtocell service offers significant competition to
macrocell service. Macrocell
operator thus has less incentive to enable femtocell service. We also investigate the issue of additional operational cost and limited coverage of femtocell service on equilibrium decisions, consumer surplus and social welfare.

\bigskip



\noindent\hrulefill

\section{Introduction}\label{sec:introduction}
Today there are over 5 billion cellphone users in the world (Global mobile
statistics 2011). Many users  experience  poor indoor
reception at home or office. This is because in the current cellular
network (also called macrocell network), high-frequency and
low-power cell signal has to travel between the outside cell site
and the indoor cell phones through various obstacles, including
brick walls, metal, and even trees, which leads to significant
signal attenuations and dropped calls (Sandler 2009).

\begin{figure}[tt]
\centering
\includegraphics[width=0.6\textwidth]{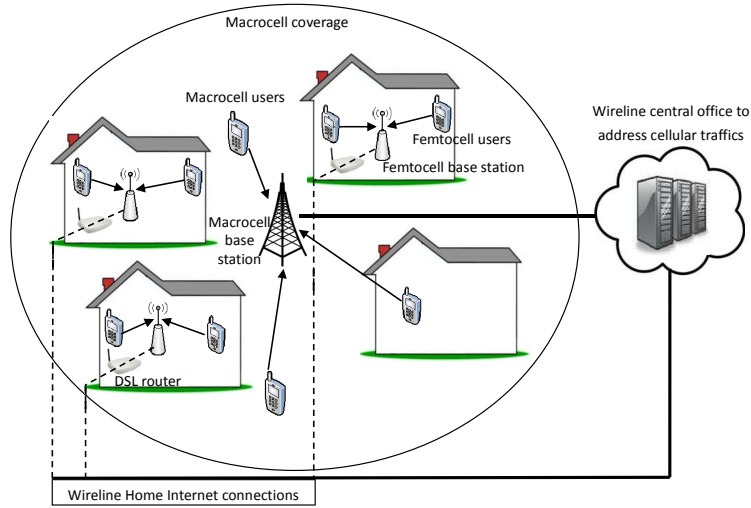}
\caption{Coexistence of femtocell service and macrocell service,
where a macrocell and three femtocells are deployed}
\label{fig:coverage_femto}
\end{figure}

One promising solution to the indoor reception
problem, femtocell technology has been rapidly developed and
deployed over the past few years. Femtocells use small base
stations of sizes similar to wireless routers. These femtocell
base stations are deployed indoors, and can pick up indoor users'
mobile signals easily and route calls to the cellular network
through home Internet connection. Femtocell technology can
significantly increase the quality of voice calls and improve the
speed of data communications (Shetty et al. 2009). Figure~\ref{fig:coverage_femto}
provides an illustration of four homes with macrocell
coverage and three of them have installed femtocell base stations.

Currently in the United States, AT\&T, Sprint Nextel and Verizon Wireless (a joint venture of Verizon Communications Inc. and Vodafone Group PLC) are already offering femtocell services to their customers. 
T-Mobile and Vodafone in Europe, NTT DoCoMo and Softbank in Japan,
and Unicom in China have been conducting tests of the technology
and planning to roll out nationwide femtocell services. In June
2010, UK research firm Informa Telecoms \& Media reported that
femtocell deployments had more than doubled in the past 12 months,
with more and more tier one operators jumping on the bandwagon
(Informa Telecoms \& Media 2011). Shipments are estimated to grow from 0.2 million units in 2009 to 12 million
units worldwide in 2014 (Berg Insight 2009).

However, one of the biggest challenges to companies' wide
 femtocell deployment is the lack of a clear business model.
As  Emin Gurdenli, chief technology officer of Deutsche Telekom AG's
T-Mobile U.K., put it  (The Wall Street Journal, Feb. 2009):

\emph{``The rationale for femtocells is well-established, but a
quantitative business case with a clear business model in terms of
how we go to market is not there yet." }

The purpose of this paper is to develop such a quantitative model to
examine the trade-off regarding femtocell deployment. In particular,
we look at the following research questions:
\begin{itemize}
\item \emph{Should current macrocell operators deploy femtocell services? How would operators allocate bandwidth (capacity) resources and make pricing decisions?} There are two common approaches to the deployment of femtocell service.
In an \emph{integrated}
system, a macrocell operator 
directly provides femtocell service to users and fully controls
bandwidth resource allocation and femtocell service price. {We also submitted a paper on the economic operation of integrated system (Duan et al. 2011).}
In a \emph{distributed} system, a macrocell operator
leases its spectrum resources  to a femtocell operator. The
femtocell operator determines the service provision and pricing independently. {We can find many such examples in industry:
Sprint leases licensed spectrum to Virgin Mobile USA to provide femtocell service (Fitchard 2009), and BT Mobile is using Vodafone's resource to provide femtocell service (Atkinson 2011). Recently there are more research on the distributed system (e.g., Hong and Tsai 2010 and Chen et al. 2011), and this paper focuses on the distributed system.}
The key tradeoff for the macrocell operator is obtaining more
revenue by leasing resources to the femtocell operator as against having
fewer resources for its own services and facing increased market
competition.
\item \emph{How would users choose between femtocell and macrocell services?} With the deployment of indoor femtocell stations, femtocell users no longer experience signal attenuation
and poor reception problem, and can achieve the maximum quality of
service. In contrast, when users are connected to the outdoor macrocell base
stations, the quality of service highly depends on the user
locations and the communication environments. When different qualities of services are coupled with different
pricing schemes, different users have different
preferences between macrocell and femtocell services.
\end{itemize}

Our main results are summarized as follows:

\begin{itemize}

\item \emph{Characterization of equilibrium decisions:} We derive the threshold of spectrum efficiency level which segments users who prefer femtocell to macrocell services. Furthermore, we characterize the femtocell operator's equilibrium femtocell price and the macrocell operator's capacity allocation and pricing decisions.
\item \emph{Analysis of impact of macrocell's total limited capacity:} Wireless spectrum is a very scarce resource so macrocell operators often face capacity
constraints. {{In the U.S. 700MHZ spectrum auction in
March 2008, the total bid price is nearly \$20b
(WNN Wi-Fi Net News 2008).}} We show that macrocell operator has more incentive to lease spectrum
to the femtocell operator when its capacity is \emph{small}, but
chooses to  offer only macrocell service when its capacity is
\emph{large}.

\item \emph{Calculation of consumer surplus and social welfare:} With no additional operational
cost and full coverage, femtocell service can increase both the
\emph{total} consumer surplus and social welfare. However, we show that some
users might experience a smaller payoff from the adoption of the
femtocell service if, for example, they do not experience much service quality improvement
with the femtocell service but need to pay a higher price.

\end{itemize}

{ In addition, we have examined two extensions of the basic model. The first is
with additional femtocell operational cost. Although femtocells
are low in \emph{deployment} costs, the femtocell service may incur
additional \emph{operational} cost compared to macrocell service.
For instance, femtocell operators may be charged by internet service
providers for routing traffic through wireline broadband internet to
reach the cellular network (McKnight et al. 1997).
The impact of the additional operator cost on the femtocell operator
is obviously negative; its impact on the macrocell operators,
however, is unclear and deserves detailed exploration. The second is the impact of
limited femtocell coverage. A femtocell base station typically has a
smaller spatial coverage. For instance,
a femtocell device may only cover a region with a diameter of 50-100 meters, 
whereas a macrocell covers a larger range with a diameter of more than
10 kilometers. {The femtocell service may have limited coverage
when it does not have enough femtocell base stations.}
We examine the impact of such limited coverage on macrocell
and femtocell operators' profits. }

The rest of the paper is organized as follows. We introduce the
network model of macrocell service in
Section~\ref{sec:NetworkModel}, which serves as a benchmark for
later analysis. In Section~\ref{sec:UnfairFemto}, we introduce the
network model of femtocell service and analyze how the macrocell
operator and femtocell operator make capacity and pricing decisions to maximize
their own profits. {Then, in Sections~\ref{sec:cost} and
\ref{sec:coverage}, we extend the results in
Section~\ref{sec:UnfairFemto}  by examining the various effects of femtocell
operational cost and limited femtocell coverage}. In Section~\ref{sec:conclusion} we present the conclusion to our study and discuss future work.

\section{Literature Review}
{ Our work is closely related to two main streams of literature: i)
studies of femtocell deployment in the telecommunication literature,
and ii) studies of dual channel competition in the management science
and operations research literature.

Most existing work on femtocell deployment in the telecommunication
literature (\eg Chandrasekhar and Andrews 2009) focus on various
technical issues in service provision such as  access control,
resource management, and interference mangement. Only a few papers
discuss the economic issues of femtocells (\eg
Claussen et al. 2007, Yun et al. 2011, Shetty et al. 2009, Chen et al. 2011), examining the
impact of network deployment costs and femtocells'
openness to macrocell users. 
The key difference between our paper and such existing literature is
that we study the provision of dual services in terms of both
spectrum allocations and pricing decisions. We also characterize the
impact of the femtocell operational cost and limited femtocell
coverage on the service provision. }

Our work is also closely related to the literature on \emph{dual
channel competition} in the area of management science and
operations research. In this body of literature, there are usually
two types of decision makers: a manufacturer and a retailer. The
manufacturer can sell the products through a direct channel, a
retailer channel, or both.
Chiang et al. study whether and how a
manufacturer should operate a new direct channel when it already has
a retailer partner. They show that direct marketing can indirectly
increase the flow of profits through a retail channel by reducing the
degree of double marginalization. Also, the direct channel may not
be a threat to the retailer since the wholesale price is driven
down.
Tsay and Agrawal 2004 further exploit several means whereby the manufacturer can
mitigate channel conflict between the direct channel and the
retailer channel, including adjustments of wholesale price, paying a
commission to a retailer, and entirely conceding demand fulfillment  on
the part of the retailer. More general results are obtained motivated by the
models in Chiang et al. 2003 and Tsay and Agrawal 2004. For example,
Huang and Swaminathan 2009 posit a stylized deterministic demand model where
each channel relies on prices, degree of substitution across
channels, and the overall market potential. Dumrongsiri et al. 2008
investigate the influence of demand variability on prices and
manufacturer's incentive to open direct channel.


In the context with which we are concerned, that of femtocell deployment, we can view the
\emph{macrocell operator} as the manufacturer, the \emph{femtocell
operator} as the retailer, and the \emph{macrocell service} as the
direct channel. Our paper has four key differences from prior
literature.

First, we consider a different order of introducing the new channel.
Instead of introducing the direct channel after the retailer channel, as is the case in Chiang et al. 2003, Tsay and Agrawal
2004a, Huang and Swaminathan 2009, Dumrongsiri
et al. 2008, we consider
the case in which the manufacturer owns the direct channel first and
decides on the best way to open the retailer channel.

Second, the limited capacity model considerably complicates the
analysis of our model. The dual channel literature generally assumes
unlimited potential supply, \ie that the manufacturer can produce as many
products as possible (with a production cost) to maximize its
profit.
However,
a macrocell operator often has only a limited total capacity in the
decision time scale considered here. This is
because the spectrum allocation to cellular service providers are
often regulated by government authorities (\eg FCC in USA and Ofcom
in UK). The macrocell operator often obtains spectrum
licenses that last for years or decades. The long license period
ensures enough motivation for the macrocell providers to invest in
the necessary network infrastructure, which is often very expensive.

Third, the heterogeneity of users in our model is motivated by the
unique characteristics of wireless communications, and is thus
different from that considered by prior literature. In particular, users have different
channel conditions (and thus different evaluations of the same
resource allocation) under the macrocell service (direct channel),
but have the same maximum channel condition under the femtocell
service (retailer channel). In contrast, prior literature either
assumes that users are homogenous or are different in willingness to
pay. 
Moreover, the users' utility functions here are also motivated by
today's wireless communication technologies, which renders some of
the prior generic analysis inapplicable.

Finally, we characterize the impact of limited femtocell coverage on
the new service provision. Very few prior studies have considered a similar
constraint. Rubin 1978 considers a related constraint where the
monitoring costs of company-owned outlets rise with physical
distance from headquarters, and thus direct channel becomes
non-profitable in suburban areas. What we considered is the
limitation of coverage of the retailer channel, and  the model thus
is different.

\section{Benchmark Scenario: Macrocell Service Only}
\label{sec:NetworkModel}

{{Throughout this paper, we focus on the monopoly case in the two-tier market with a single macrocell operator and a single femtocell operator. The reason is that we can observe many monopoly examples in macrocell service worldwide (e.g., America Movil (the world's fourth largest mobile network operator in Mexico and many places in Latin America), and MTS in some central Asian countries). Also, since femtocell service just emerged from last decade, many fetmocell operators are still local monopolists (e.g., Virgin Mobile USA in US and BT Mobile in UK). Moreover, monopoly is a first step leading to a more general oligopoly market and we plan to study oligopoly case as a future direction as in Section~\ref{sec:conclusion}.}}

As a benchmark case, we first look at how the macrocell operator
prices the macrocell service to maximize its profit without
introducing the femtocell service.
When we consider the introduction of femtocell service in Sections
\ref{sec:UnfairFemto} and \ref{sec:cost}, the macrocell operator
needs to achieve a profit no worse than this benchmark case.


For the sake of discussion, we will focus on the operation of a
single macrocell. {In general, a macrocell operator owns
multiple macrocells. Non-adjacent macrocells can share the same
frequency (called frequency reuse). The analysis of this paper can
be extended to the more general case without changing the main
managerial insights.}

The macrocell operator owns wireless spectrum
(also called bandwidth) with a limited capacity; a user needs to
access the bandwidth in order to complete its wireless
communications (\eg voice calls, video streaming, data transfer). A
larger bandwidth means more resources to the user and thus better
communication quality of service (QoS), but also leads to a greater
expense.

\begin{figure}[tt]
\centering
\includegraphics[width=0.8\textwidth]{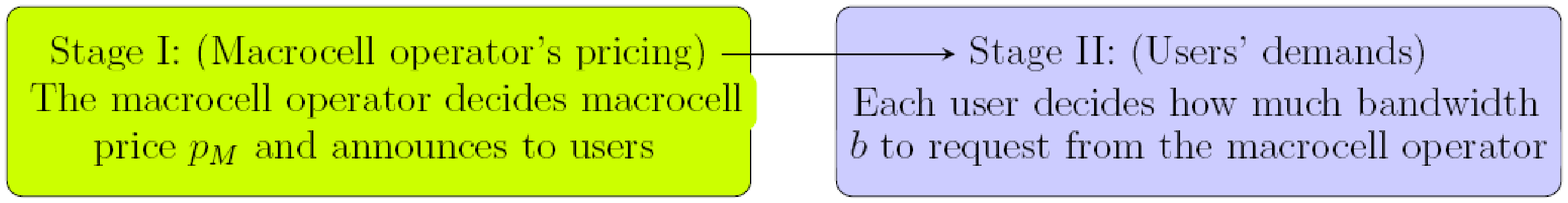}
\caption{Two-stage Stackelberg game between the macrocell operator and users.}
    \label{fig:twostage}
\end{figure}

As shown in Fig.~\ref{fig:twostage},  we model the interactions
between the macrocell operator and end users as a two-stage
Stackerberg game. In Stage I, the macrocell operator determines the
macrocell price $p_M$ per unit bandwidth.
In Stage II, each user decides how much bandwidth to purchase. The operator wants to maximizes its profit,
while the users want to maximize their payoff.
Such usage-based pricing scheme is widely used in
today's cellular macrocell networks, especially in Europe
and Asia (Courcoubetis and Weber 2003, Altmann and Chu 2001). In US, AT\&T (since a year ago) and
Verizon (since July 2011) have adopted the usage-based
pricing for wireless data services. Usage-based pricing
for femtocells has just started. For example, AT\&T's femtocell
service counts the femtocell data usage as part of
the regular cellular usage (together with the macrocell
data usage), which is subject to usage-based pricing
(AT\&T 2011). Due to the exponential growth of wireless data
traffic and the scarce spectrum resource, we envision that
usage-based pricing for both macrocell and femtocell
services will become more and more common in the near future.


Next, we solve this two-stage Stackelberg game by backward induction
(Myerson 1997).

\begin{figure}
\unitlength=1mm
\begin{picture}(90, 15)
\put(6,5){\huge\line(1,0){70}} \put(73.2,3.97){$\rightarrow$}
\put(6,4){\line(0,1){2}}  \put(30,4){\line(0,1){2}}
\put(73.3,4){\line(0,1){2}} \put(9.7,6){No service}
\put(38,6){Macrocell service} \put(5.2,1){$0$} \put(28,1){$p_M$}
\put(73,1){$1$} \put(80,4){Macrocell spectrum efficiency $\theta$}
\end{picture}
\caption{Distribution of users' macrocell spectrum efficiency
$\theta$} \label{fig:User_theta}
\end{figure}
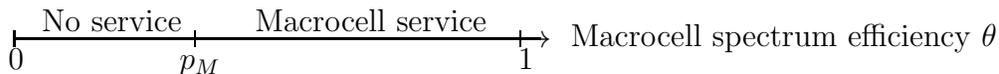

%
%

\subsection{Stage II: Users' Bandwidth Demand Given A Fixed Price $p_{M}$}

The QoS of a wireless communication session depends not only on the
resource allocation but also on the condition of the wireless
channel between the transmitter and receiver. The channel condition
is determined by both the locations of transmitter and receiver and
the surrounding environment. As an example, let us consider uplink
transmissions from the users mobile phones to the common single
macrocell base station (as in Fig.~\ref{fig:coverage_femto}). The
channel condition in general decreases with the distance between the
user and the base station, and can become very weak if the user is
inside a house with thick walls. A user with a bad channel condition
will not be able to achieve a high data rate even with a large
bandwidth allocation.

Here we model the users' channel heterogeneity by a \emph{macrocell
spectrum efficiency} $\theta$, which is assumed to be uniformly
distributed in $[0,1]$ (see Fig.~\ref{fig:User_theta}). {The
uniform distribution is assumed for analytical tractability. A more
complicated distribution based on field measurements will not change
the main managerial insights obtained in this paper.}
%
%
%
A larger $\theta$ means a better channel condition and a higher
spectrum efficiency \emph{when using the macrocell service}.

For a user with a macrocell spectrum efficiency $\theta<1$, when
allocated macrocell bandwidth $b$, its effective resource allocation
is $\theta b$. Its \emph{utility} $u(\theta,b)$ (\eg data rate) can
be modeled as ({similar to} Sengupta and Chatterjee 2009, Wang and Li 2005)
$$u(\theta,b)=\ln(1+\theta b),$$
which is concave in $b$ representing the diminishing return in
bandwidth consumption. {The more bandwidth a user obtains, he can experience a higher data rate and a better QoS when communicating with others.}
The user needs to pay a linear payment $p_Mb$ to the macrocell
operator, where the price $p_M$ is announced by the macrocell
operator in Stage I. {{Note that the usage-based pricing
is becoming a main trend in macrocell service market (and replacing
flat-fee pricing for data traffics) (Goldstein 2011).}} The user's
\emph{payoff} is the difference between the utility and payment, \ie
\begin{equation}\label{eq:payoff_M}
r_{M}(\theta, b,p_{M})=\ln(1+\theta b)-p_Mb.
\end{equation}

The optimal value of bandwidth (demand) that maximizes the user's
payoff with the macrocell service is
{\begin{equation}\label{eq:demand_M}
b^{\ast}(\theta,p_{M})= \frac{1}{p_M}-\frac{1}{\theta}
\end{equation}
if $p_M \leq \theta$ and $0$ otherwise.}
Notice that $b^{\ast}(\theta,p_{M})$ is decreasing in $p_M$ and increasing in $\theta$ (if $p_M
\leq \theta$). When $p_{M}>\theta$, the user chooses not to start
the wireless communication as it is too expensive (by taking its
macrocell spectrum efficiency $\theta$ into consideration).
{The user's maximum payoff with macrocell service is
\begin{equation}\label{eq:payoff_M_opt}
r_{M}(\theta, b^{\ast}(\theta,p_{M}),p_{M})=\ln\left(\frac{\theta}{p_M}\right)-1+\frac{p_M}{\theta}
\end{equation}
if $p_M \leq \theta$ and 0 otherwise.}
Notice that the payoff is always nonnegative.

\subsection{Stage I: Macrocell Operator's Pricing $p_{M}$}

Next we consider the macrocell operator's optimal choice of price
$p_{M}$ in Stage I. To achieve a positive profit, the macrocell
operator needs to set $p_{M}\leq \max_{\theta\in[0,1]}\theta=1$,
otherwise no user will request any bandwidth in Stage II.

Without loss of generality, we normalize the total user population
to 1. The fraction of users choosing macrocell service is $1-p_M$
as shown in Fig.~\ref{fig:User_theta}.  The total user demand is
\begin{equation}
Q_M(p_M)=\int_{p_M}^1\left(\frac{1}{p_M}-\frac{1}{\theta}\right)d\theta=\frac{1}{p_M}-1+\ln
p_M,
\end{equation}
which is a decreasing function of $p_M$.

Recall that the macrocell operator has a limited bandwidth capacity
$B$, and thus can only satisfy a total demand no larger than $B$.
The macrocell operator's profit is
\[\pi^{macro}(p_M)=p_M\min\left(B,\frac{1}{p_M}-1+\ln p_M\right).
\]
The operator will choose price $p_{M}$ to maximize profit, \ie
\begin{equation}\label{eq:macroprofitmax}
\max_{0<p_M\leq 1}\pi^{macro}(p_M).
\end{equation}

Theorem \ref{thm:Direct_optimal} characterizes the unique optimal
solution to Problem (\ref{eq:macroprofitmax}).

\begin{theorem}\label{thm:Direct_optimal}
The equilibrium macrocell price $p_M^{bench}$ that maximizes the
macrocell operator's profit  in the two-stage Stackelberg game in
Fig.~\ref{fig:twostage} is the unique solution to the following
equation:
\begin{equation}\label{eq:p_M_opt}
B=\frac{1}{p_{M}}-1+\ln p_{M}.
\end{equation}
The total user demand equals the maximum capacity at the
equilibrium, \ie $Q_M(p_{M}^{bench})=B$. The equilibrium price
$p_{M}^{bench}$ decreases with $B$, and the macrocell operator's
equilibrium profit $\pi^{macro}(p_M^{bench})$ increases with $B$.
\end{theorem}

Notice that no users with a macrocell spectrum efficiency $\theta$
less than  $p_{M}^{bench}$ will  receive macrocell service. When the
total bandwidth $B$ is small, the equilibrium macrocell price
$p_{M}^{bench}$ is close to 1 and thus most users will not get
service. This motivates the macrocell operator to adopt the
femtocell service, which is able to serve these users and leads to
additional profit, which is shared by the macrocell and femtocell operators.

\section{Femtocell Deployment}\label{sec:UnfairFemto}
We now turn to the question of how femtocell service may improve the macrocell
operator's profit.
%
We are interested in understanding the following issues:
\begin{itemize}
\item \emph{Strategic decision:} Is it economically beneficial for the  macrocell operator to lease spectrum to a femtocell operator, who will compete with the macrocell operator in serving the same group of end users?
\item \emph{Operational decisions:} If the answer to the previous question is yes, how should the macrocell operator allocate and price the spectrum resources?
\end{itemize}

The analysis in this section is based on several simplified
assumptions:
\begin{itemize}
\item The femtocell service does not incur any additional operational cost compared to the macrocell service.
This assumption will be relaxed in Section~\ref{sec:cost}.
\item The femtocell service has the same coverage as the macrocell service,
such that each user has the choice between two services. This
assumption will be relaxed in Section~\ref{sec:coverage}.
\end{itemize}

{It should be noted that if we simultaneously relax both the two assumptions later, it is difficult to see the impact of each explicitly. Thus we relax them in separate sections.}

\begin{figure}[tt]
\centering
\includegraphics[width=0.9\textwidth]{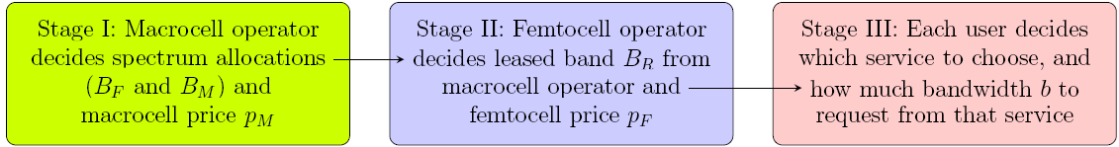}
\caption{Three-stage dynamic game between the macrocell operator, femtocell operator, and users.}
    \label{fig:threestage_new}
\end{figure}

More specifically, we will look at a three-stage dynamic game as in
Fig.~\ref{fig:threestage_new}. {The macrocell operator has market power and is the leader in the cellular market, while the emerging femtocell operaotr is the follower}. In Stage I, the macrocell operator
decides bandwidth allocations to femtocell service $B_F$ and
macrocell service $B_M$ such that $B_F+B_M=B$. {{Here we focus on ``separate carriers'' scheme where dual
services operate on independent spectrum bands. ``Separate carriers'' is easy to manage and
can avoid interferences between macrocells and femtocells. For example, China Unicom (one of
the top 3 wireless service providers in China and the first one deploying femtocell since
2009) is in strong favor of this scheme (China Femtocell Symposium 2011). There exists another scheme called ``shared carriers'' where
dual services operate on the same spectrum bands, which is discussed in Section~\ref{sec:conclusion}.}} The macrocell operator also decides the
macrocell price $p_{M}$, which is charged to both the femtocell
operator and end users who choose macrocell services. {Note that we assume
the same price is charged to both femtocell operator and users of
macrocell services, to avoid arbitrage opportunity.} {For example, if the macrocell operator charges the femtocell operator more
than macrocell users, the femtocell operator can pretend to be
macrocell users, request spectrum at the macrocell price, and serve its femtocell users. Such an approach has been studied in
Tanneur 2003. If the macrocell operator charges the
femtocell operator less than macrocell users, then some other
intermediate operator, e.g., mobile virtual
network operators, can  disguise themselves as
femtocell operators and obtain spectrum resource at a lower price
than macrocell price, and then provide macrocell service to users to make a profit. This eventually decreases the
macrocell operator's market share and profit. Such scenarios have been considered in Dewenter and Haucap 2006. }

In Stage II, femtocell operator decides how much bandwidth $B_R$ to
lease from the macrocell operator such that $B_R\leq B_F$, and pay
$B_Rp_M$ to the macrocell operator.
It also determines femtocell price $p_F$ to end users choosing the
femtocell service. In Stage III, each user decides which service to
choose and how much bandwith to purchase for the service. If a
user's preferred service is not available (when demand is larger
than capacity for that service), the user will seek the other
service. {Here we focus on a large group of users, where a single
user's demand is infinitesimal to the total demand. Thus we can
ignore cases in which a user purchases bandwidth from both
services.}

{Our considered decision process in Stages I and II is like the
existing spectrum auction, where the spectrum holder announces total bandwidth amount and price first,
then bidders request and give out payment. There may possibly exist some other decision process, e.g., femtocell operator
requests some bandwidth from macrocell operator first and then macrocell operator decides how much to satisfy at some price.
Intuitively, these two processes should lead to the same equilibrium outcome. This is because the macrocell operator can adjust macrocell price to
make its decided femtocell band matches femtocell operator's demand. It should also be noted that it is optimal for the femtocell operator to announce femtocell price after purchasing bandwidth from the macrocell operator. Otherwise, it may run out of bandwidth at a low price or waste some bandwidth.}

We will again analyze this three-stage dynamic game using
backward induction.

To differentiate from the macrocell service only benchmark in
Section \ref{sec:NetworkModel}, we refer to the setup in this
section as \emph{dual services}. Notice that dual services may
degenerate to the case of benchmark when the macrocell operator
decides not to lease spectrum to the femtocell operator, \ie when
the total capacity is large as shown later in this section.

\subsection{Stage III: Users' Service Choice and Bandwidth Demand}\label{subsec:StageII_dual}

If a user has a macrocell spectrum efficiency $\theta$, his optimal
payoff by using the macrocell service is  given in
(\ref{eq:payoff_M_opt}).
Next, we consider a user's payoff by using the femtocell service.

Since femtocell base stations are deployed indoors and are very
close to users' cell phones, it is reasonable to assume that all
users using the femtocell service have equally good channel conditions
and achieve the same \emph{maximum} spectrum efficiency. This means
that, independent of the macrocell spectrum efficiency $\theta$, each
user achieves the same \emph{payoff} $r_{F}(b,p_F)$ when using a
bandwidth of $b$ under femtocell service,
%
\begin{equation}\label{eq:payoff_Femto}
r_{F}(b,p_{F})=\ln(1+b)-p_Fb.
\end{equation}
The user's optimal demand in femtocells is
\begin{equation}\label{eq:femtodemand}
b^{\ast}(p_{F})= \frac{1}{p_F}-1
\end{equation}
if $p_F \leq 1$ and 0 otherwise.
A user's maximum payoff under the femtocell service is
\begin{equation}\label{eq:payoff_Femto_opt}
r_{F}\left(b^{\ast}(p_{F}),p_{F}\right)=\ln\left(\frac{1}{p_F}\right)-1+p_F
\end{equation}
{if $p_F \leq 1$,} and $0$ otherwise.
which is always nonnegative.
Note that {{some
operators have adopted a flat-fee charging scheme for femtocell
services to encourage early user adoptions. In this paper, we focus
on analyzing the usage-based pricing for femtocell services in a
mature market. It should also be noted that usage-based pricing
often leads to a higher profit than the flat-fee charging
(Courcoubetis and Weber 2003).}}

We will show that $p_{F}>p_{M}$ at the equilibrium, \ie the
femtocell price $p_{F}$ in Stage II, is always larger than the
macrocell price $p_{M}$ in Stage I. By comparing the user's payoffs
in (\ref{eq:payoff_M_opt}) and (\ref{eq:payoff_Femto_opt}), it is
clear that a user with $\theta=1$ will always
choose macrocell service to maximize his payoff. On the other hand,
a user with a small $\theta$ would choose femtocell service to
improve his payoff. As a result, we define the following thresholds
of $\theta$ that separate the user population into two service
groups.

\begin{definition}[Users' preferred partition threshold $\theta_{th}^{pr}$]
Users with $\theta\in[0,\theta_{th}^{pr})$ prefer to use the
femtocell service, and users with
$\theta\in[\theta_{th}^{pr},1]$ prefer to use the macrocell service. 
\end{definition}

\begin{definition}[Users' finalized partition threshold $\theta_{th}$] \label{def:partthrehold}
The finalized partition threshold $\theta_{th}$ is the minimum
macrocell spectrum efficiency among all the users actually served by
the macrocell service. Users with $\theta\in[\theta_{th},1]$ receive
the macrocell service, while users with $\theta\in[0,\theta_{th})$
receive either the femtocell service or no service.
\end{definition}

The preferred partition threshold $\theta_{th}^{pr}$ only depends on
prices $p_{M}$ and $p_{F}$.
If all users' demands from their preferred services are satisfied,
then users' preferred partition threshold equals users' partition
threshold (\ie $\theta_{th}^{pr}=\theta_{th}$). However, in general
$\theta_{th}$ may be different from $\theta_{th}^{pr}$, depending on
the values of $B_F$, $B_M$, and $B_R$ in Stages I and II.

By comparing a user's optimal payoff with macrocell and femtocell
services in (\ref{eq:payoff_M_opt}) and (\ref{eq:payoff_Femto_opt}),
we obtain the following result.


\begin{lemma}
Users' preferred partition threshold $\theta_{th}^{pr}={p_M}/{p_F}$.
\end{lemma}

Now we introduce the concept of \emph{finalized demand}.

\begin{definition}[User's Finalized Demand]
If a user's demand from his preferred service is satisfied, then his
finalized demand equals his preferred demand. If not, the user
switches to the alternative service and the new demand becomes the
finalized demand.
\end{definition}

Note that a user's finalized demand may not be realized, \eg when
the price is set too low and the total finalized demand
is larger than the supply.

\subsection{Stage II: Femtocell Operator's Spectrum Purchase and Pricing}

Now we analyze Stage II, where the femtocell operator determines
$B_R$ and $p_F$ to maximize its profit. In this stage, the macrocell
operator's decisions on $p_M$ and $B_F$ (and $B_M=B-B_F$) are
determined and  known to the femtocell operator. Let us denote the
femtocell operator's equilibrium decisions as $B_R^\ast(p_M,B_F)$
and $p_F^\ast(p_M,B_F)$, both of which are functions of $p_M$ and
$B_F$.

To maximize profit, the femtocell operator
needs to know which users will choose femtocell service and their
characteristics. Users with macrocell spectrum efficiency
$\theta\in[0,\theta_{th}^{pr})=[0,p_{M}/p_{F})$ will choose
femtocell service first. Some other users may also choose femtocell
service if their demands cannot be satisfied by the macrocell
services. The following lemma, however, shows that the macrocell
operator will reserve enough bandwidth $B_{M}$ during Stage I, such
that all users who prefer to use macrocell service will be able to
do so.
\begin{lemma}\label{lemma:no_switchers}
At the equilibrium of the three-stage dynamic game as in
Fig.~\ref{fig:threestage_new}, the macrocell operator satisfies all
preferred demands from users with
$\theta\in[\theta_{th}^{pr},1]=[p_{M}/p_{F},1]$ in macrocell
service.
\end{lemma}

{Lemma~\ref{lemma:no_switchers} is derived regardless of the decisions of femtocell operator. Thus at the equilibrium of the whole three-stage game, Lemma~\ref{lemma:no_switchers} holds and we can use it to derive femtocell operator's equilibrium decisions in Stage II.}

We will further discuss the intuitions of
Lemma~\ref{lemma:no_switchers} in the next subsection. Since
femtocell operator  only serves users with
$\theta\in[0,{p_M}/{p_F})$, its profit is
\begin{align}\label{eq:femtocellProfit}
\pi^{Femto}(p_F,B_R)&=p_F\min\left(B_R,\int_0^{\frac{p_M}{p_F}}\left(\frac{1}{p_F}-1\right)d\theta\right)-p_MB_R\nonumber\\
&=\min\left((p_F-p_M)B_R,(1-p_F)\frac{p_M}{p_F}-p_MB_R\right).
\end{align}
The femtocell operator's profit-maximization problem is:
\begin{align}\label{eq:opt_Distributed_Femto}
&\max_{p_F\geq 0,B_R\geq 0}\pi^{Femto}(p_F,B_R)\nonumber\\
&\text{subject to    }  B_R\leq B_F.
\end{align}
By solving Problem~(\ref{eq:opt_Distributed_Femto}), we have the
following result.

\begin{lemma}\label{lemma:femto}
In Stage II, the femtocell operator's equilibrium femtocell price is
\begin{equation}\label{eq:pF_Femto}
p_{F}^{\ast}(p_M,B_F)=\max\left(\frac{2p_M}{1+p_M},\frac{-p_M+\sqrt{(p_M)^2+4B_Fp_M}}{2B_F}\right),
\end{equation}
and its equilibrium femtocell bandwidth purchase is
\begin{equation}\label{eq:BR_Femto}
B_{R}^{\ast}(p_M,B_F)=\min\left(\frac{1-(p_M)^2}{4p_M},B_F\right),
\end{equation}
which equals users' total preferred demand in femtocell service.
Then users' preferred partition threshold equals equilibrium
partition threshold (\ie {$\theta_{th}^{pr}=\theta_{th}$}).
\end{lemma}

The proof of Lemma~\ref{lemma:femto} is given in Appendix~1. Lemma~\ref{lemma:femto} shows that
the femtocell operator will also satisfy the users' preferred
demands, and it does not want the users to switch to its competitor
(\ie the macrocell operator).

\subsection{Macrocell Operator's Spectrum Allocations and Pricing
in Stage I}\label{subsec:StageIII}

Now we come back to Stage I, where the macrocell operator determines
$p_M$, $B_F$, and $B_M$ to maximize its profit. Let us denote the
macrocell operator's equilibrium decisions as $p_M^\ast$,
$B_F^\ast$, and $B_M^\ast$.

Notice that Lemma~\ref{lemma:no_switchers} shows that it is optimal
for the macrocell operator to serve all users with
$\theta\in[{p_M}/{p_{F}^\ast(p_M,B_F)},1]$ by macrocell service,
where $p_{F}^\ast(p_M,B_F)$ is the equilibrium femtocell price given
in Lemma~\ref{lemma:femto}. This means that the macrocell operator
does not want  users with large macrocell spectrum efficiency
$\theta$ to choose its competitor (\ie the femtocell operator).
Intuitively,
users with a large $\theta$ demand more bandwidth in macrocell
service than in femtocell service, and thus lead to a larger profit to
the macrocell operator if they choose macrocell service.

Since $B_M=B-B_F$, we can write the macrocell operator's profit as a
function of $p_{M}$ and $B_{F}$, \ie
\begin{equation}\label{eq:macrocellProfit}
\pi^{Macro}(p_M,B_F)=p_MB_{R}^\ast(p_M,B_F)+p_M\int_{\frac{p_M}{p_{F}^\ast(p_M,B_F)}}^1\left(\frac{1}{p_M}-\frac{1}{\theta}\right)d\theta.
\end{equation}
The macrocell operator's profit-maximization problem is
\begin{align}\label{eq:opt_Distributed_Macro}
&\max_{p_M\geq 0,B_F\geq 0}\ \ \ \ \ \pi^{Macro}(p_M,B_F)\nonumber\\
&\text{subject to}\ \ \ \ \
\int_{\frac{p_M}{p_{F}^\ast(p_M,B_F)}}^1\left(\frac{1}{p_M}-\frac{1}{\theta}\right)d\theta\leq
B-B_F,
\end{align}
where $p_{F}^\ast(p_M,B_F)$ and $B_{R}^\ast(p_M,B_F)$ are given in
(\ref{eq:pF_Femto}) and (\ref{eq:BR_Femto}), respectively. The
constraint shows that macrocell band $B_M=B-B_F$ can satisfy
 users' total preferred macrocell demand.

Problem~(\ref{eq:opt_Distributed_Macro}) is not convex and is
difficult to solve in closed-form, but can be solved easily using
numerical methods. Next we introduce a useful lemma that facilitates
our later discussions on numerical results.

The macrocell operator wants to sell its total capacity $B$ at the
highest macrocell price $p_M$. Under a fixed price, the total demand
from the users depends on which services they subscribe to.
If we can maximize the user demand under any fixed
price, then we can achieve the maximum revenue by optimizing the
choice of price accordingly.

Recall that a user's demand is $\frac{1}{p_F^\ast(p_M,B_F)}-1$ in
femtocell service and  $\frac{1}{p_M}-\frac{1}{\theta}$ in macrocell
service. We have the following lemma.
\begin{lemma}\label{lemma:partition}
The macrocell operator wants users with
$\theta\in\left[0,\frac{1}{\frac{1}{p_M}-\frac{1}{p_F^\ast(p_M,B_F)}+1}\right)$
to choose femtocell service, and rest of the users with
$\theta\in\left[\frac{1}{\frac{1}{p_M}-\frac{1}{p_F^\ast(p_M,B_F)}+1},1\right]$
to choose macrocell service. That is, it prefers users' partition
threshold to be
$\tilde{\theta}_{th}=\frac{1}{\frac{1}{p_M}-\frac{1}{p_F^\ast(p_M,B_F)}+1}$.
\end{lemma}

Note that the threshold in Lemma \ref{lemma:partition} is what the
macrocell operator wants to see; however, it may not equal the
users' finalized partition threshold
$\theta_{th}=\frac{p_M}{p_F^\ast(p_M,B_F)}$. This is because that
the macrocell operator cannot fully control the femtocell operator's
decisions. The difference between these two thresholds are due to
the market competition between macrocell and femtocell operators.

\subsection{Numerical Results}\label{sec:unfairfemtollnum}

Solving Problem~(\ref{eq:opt_Distributed_Macro}) numerically, we
obtain the macrocell operator's equilibrium femtocell band
$B_F^\ast$, and macrocell band $B_M^\ast=B-B_{F}^{\ast}$, and the
macrocell price $p_{M}^{\ast}$. Plugging into (\ref{eq:pF_Femto}), we
obtain the equilibrium femtocell price $p_{F}^{\ast}$.

\begin{figure*}
   \begin{minipage}[t]{0.49\linewidth}
      \centering
      \includegraphics[width=1\textwidth]{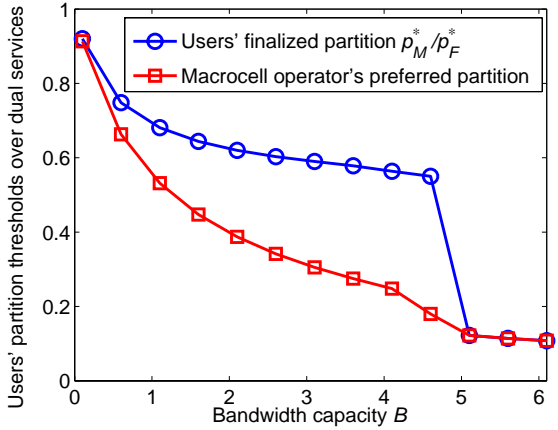}
      \caption{Users' equilibrium partition threshold
$\theta_{th}=\frac{p_M^\ast}{p_F^\ast}$ and macrocell operator's
preferred
$\tilde{\theta}_{th}=\frac{1}{\frac{1}{p_M^\ast}-\frac{1}{p_F^\ast}+1}$
as functions of $B$}
\label{fig:unfair_distributed_partition_B}
   \end{minipage}
   \begin{minipage}[t]{0.49\linewidth}
      \centering
      \includegraphics[width=1\textwidth]{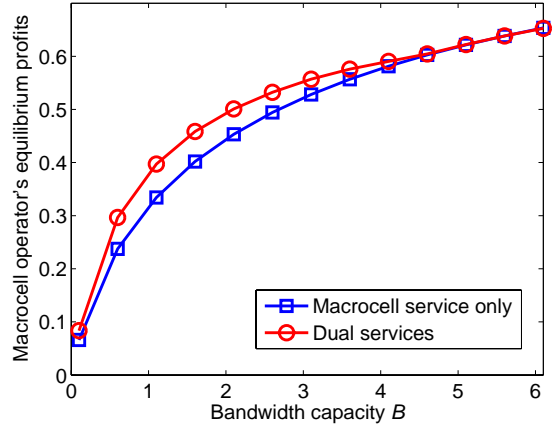}
      \caption{Macrocell operator's profits in macrocell service only case
and current dual service case as functions of capacity $B$}
\label{fig:unfair_distributed_profit_B}
   \end{minipage}
\end{figure*}

{ Figure~\ref{fig:unfair_distributed_bands_B} shows the macrocell
operator's equilibrium bandwidth allocation. The X axis is total
bandwidth capacity $B$ and the Y axis is the macrocell and femtocell
bandwidth $B_F^\ast$ and $B_M^\ast$, respectively. It shows
that when the total bandwidth capacity $B$ is small, the macrocell
operator would lease spectrum to the femtocell operator, so both
macrocell and femtocell services are provided to end users; however,
when the total bandwidth capacity $B$ becomes large, only macrocell
service is provided (i.e, $B_F^\ast=0$ and $B_M^\ast=B$). The intuition behind this is as follows: with large bandwidth capacity, the
macrocell operator can already serve most users by macrocell
service. The potential benefit of reaching the remaining small
portion of customers through facilitating femtocell service can not
compensate the potential loss due to new market competition. Hence
the  macrocell operator has no motive to lease spectrum to femtocell
provider. However, with small capacity $B$, the macrocell service
price $p_M^\ast$ is high, and thus most users with
$\theta\in[0,p_M^\ast)$ would not request macrocell service. By
leasing bandwidth to femtocell operator, the macrocell operator can
obtain a larger profit from serving more users (indirectly through
femtocell operator). {It should be noted that $B=4.77$ (that distinguishes total capacity to be small or large) is a normalized value compared to users' population, as we have normalized users' population to be $1$.}

\begin{figure*}
   \begin{minipage}[t]{0.49\linewidth}
      \centering
      \includegraphics[width=1\textwidth]{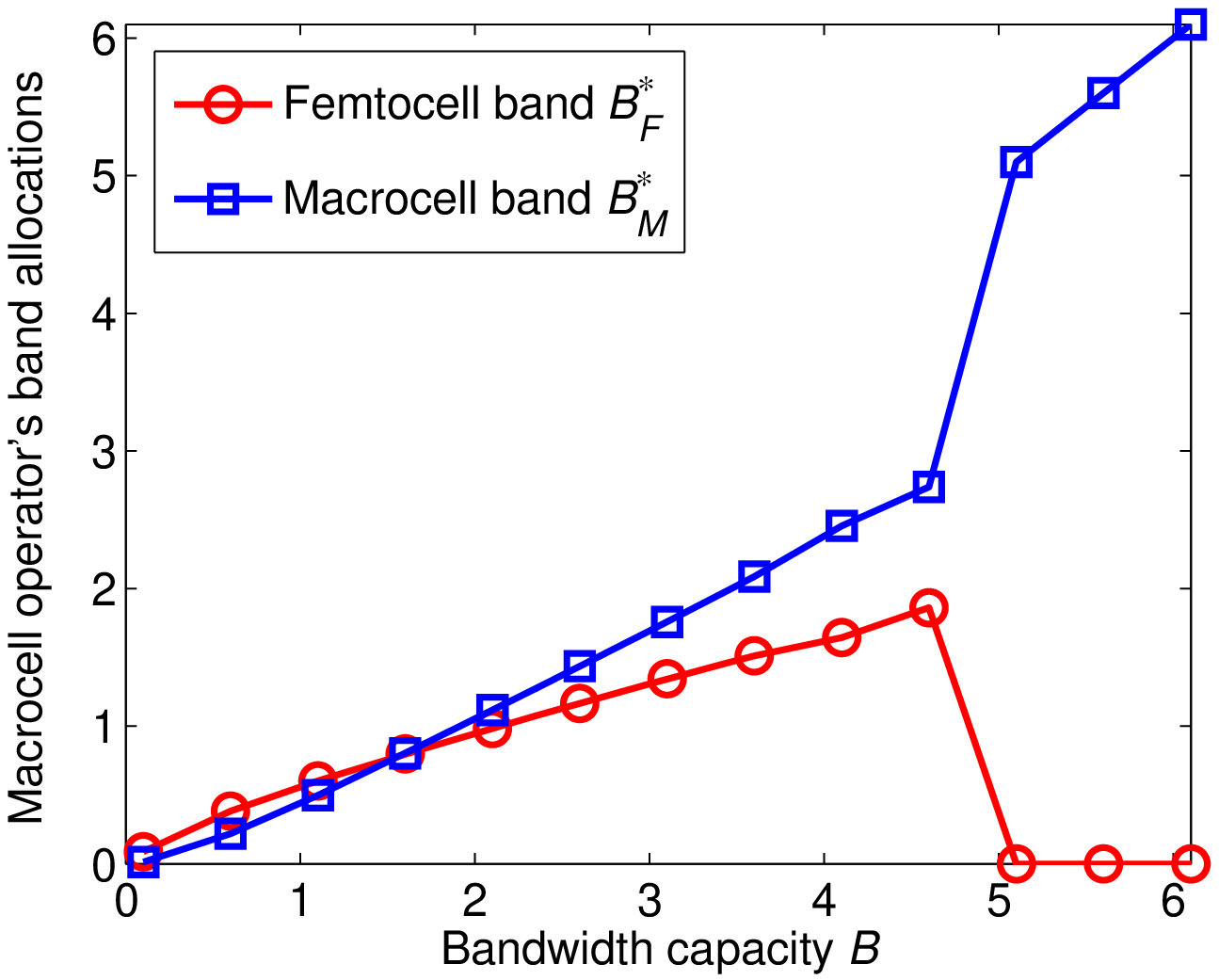}
      \caption{Equilibrium bands
      $B_{F}^{\ast}$ and $B_{M}^{\ast}$ as functions of
      capacity $B$} \label{fig:unfair_distributed_bands_B}
   \end{minipage}
   \begin{minipage}[t]{0.49\linewidth}
      \centering
      \includegraphics[width=1\textwidth]{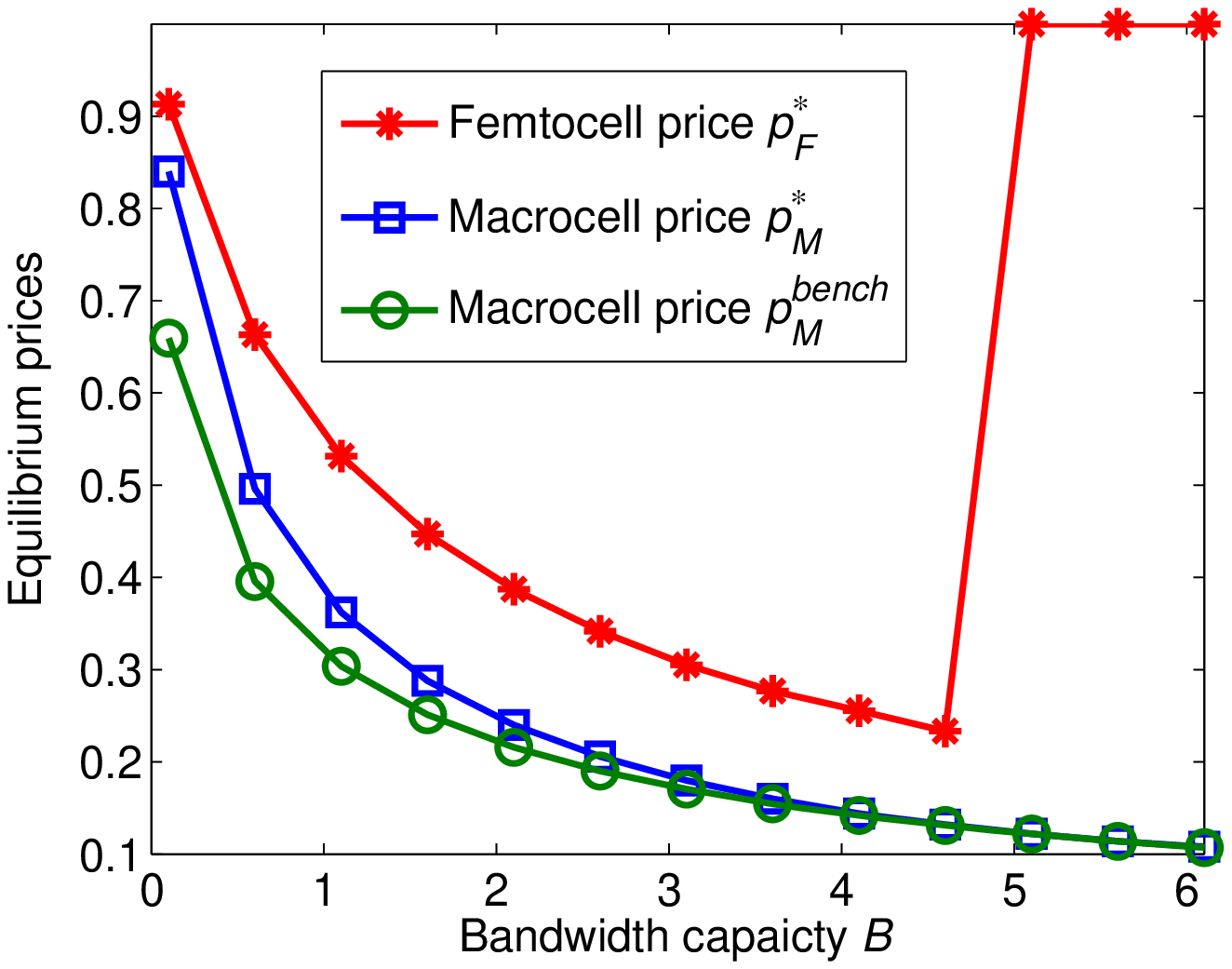}
      \caption{$p_{F}^{\ast}$ and $p_{M}^{\ast}$ in dual services, and benchmark price
       $p_M^{bench}$ as functions of $B$}
      \label{fig:unfair_distributed_prices_B}
   \end{minipage}
\end{figure*}

Figure~\ref{fig:unfair_distributed_prices_B} shows how the
femtocell and macrocell prices $p_{F}^{\ast}$ and $p_{M}^{\ast}$
change in the total bandwidth capacity. It also shows the macrocell
price of the benchmark case $p_M^{bench}$  where there is macrocell
service only. First, we observe that when the total bandwidth
capacity $B$ becomes large, the femtocell price $p_{F}^{\ast}$
becomes 1 and $p_{M}^{\ast}=p_M^{bench}$. This essentially means
that only macrocell service is provided, which is consistent with
the observation from Fig.~\ref{fig:unfair_distributed_bands_B}.
Second, the equilibrium macrocell price $p_M^\ast$ is always no less
than the benchmark price $p_M^{bench}$. This means that the
macrocell operator can obtain a larger profit with femtocell deployment by reaching more users.

Figure~\ref{fig:unfair_distributed_partition_B} shows users'
finalized partition threshold
$\theta_{th}=\frac{p_M^\ast}{p_F^\ast}$, and compares with the
threshold
$\tilde{\theta}_{th}=\frac{1}{\frac{1}{p_M^\ast}-\frac{1}{p_F^\ast}+1}$
that macrocell operator prefers. It shows that the gap
increases in the total capacity $B$. This means that femtocell operator
attracts more original macrocell users to femtocell service, and
competition between two operators becomes more intense as $B$
increases.

}

{
By summarizing the results in Figs.~\ref{fig:unfair_distributed_bands_B} to \ref{fig:unfair_distributed_partition_B}, we have the following result.

\begin{observation}\label{ob:interesting}
Only when its total bandwidth capacity $B$ is small, the macrocell operator will lease spectrum to the femtocell operator to provide femtocell service and thus serve more users. When $B$ is large, the macrocell operator will not lease any spectrum to eliminate significant competition from femtocell operator.
\end{observation}

Today's macrocell operators have ``small'' capacities in big cities, in the sense
that their capacities are not enough to match the too many people's sharp growth of wireless data demand. For example, we can witness poor service coverage of AT\&T's macrocell service in New York City and San Francisco due to lack of spectrum resource (LaVallee 2009, WNN Wi-Fi Net News 2008). The absolute value of total capacity is small and we have small $B\leq 4.77$ in these cities (WNN Wi-Fi Net News 2008). Thus we can observe macrocell operators' strong incentives to lease spectrum to femtocell operators in big cities(e.g., Sprint to Virgin Mobile USA and Vodafone to BT Mobile) to serve more users. However, in many other places with fewer user density, femtocell services haven't been deployed yet.}

{
Next we investigate how the introduction of femtocell service
affects the macrocell operator's profit, consumer surplus (\ie users' aggregate payoff), and the
social welfare (\ie summation of the profits of both operators and
the payoffs of all users). In each figure, we compare the dual
services with the macrocell service only benchmark. Our discussions
focus on the dual-service region.

Figure~\ref{fig:unfair_distributed_profit_B} shows the profits
of the macrocell operator when both services are provided vs. when
only macrocell service is provided, {both of which are
increasing in capacity $B$.
The
provision of femtocell service improves the macrocell operator's
profit significantly, especially when $B$ is small (e.g., $27\%$
when $B=0.1$).}

\begin{figure*}
   \begin{minipage}[t]{0.49\linewidth}
      \centering
      \includegraphics[width=1\textwidth]{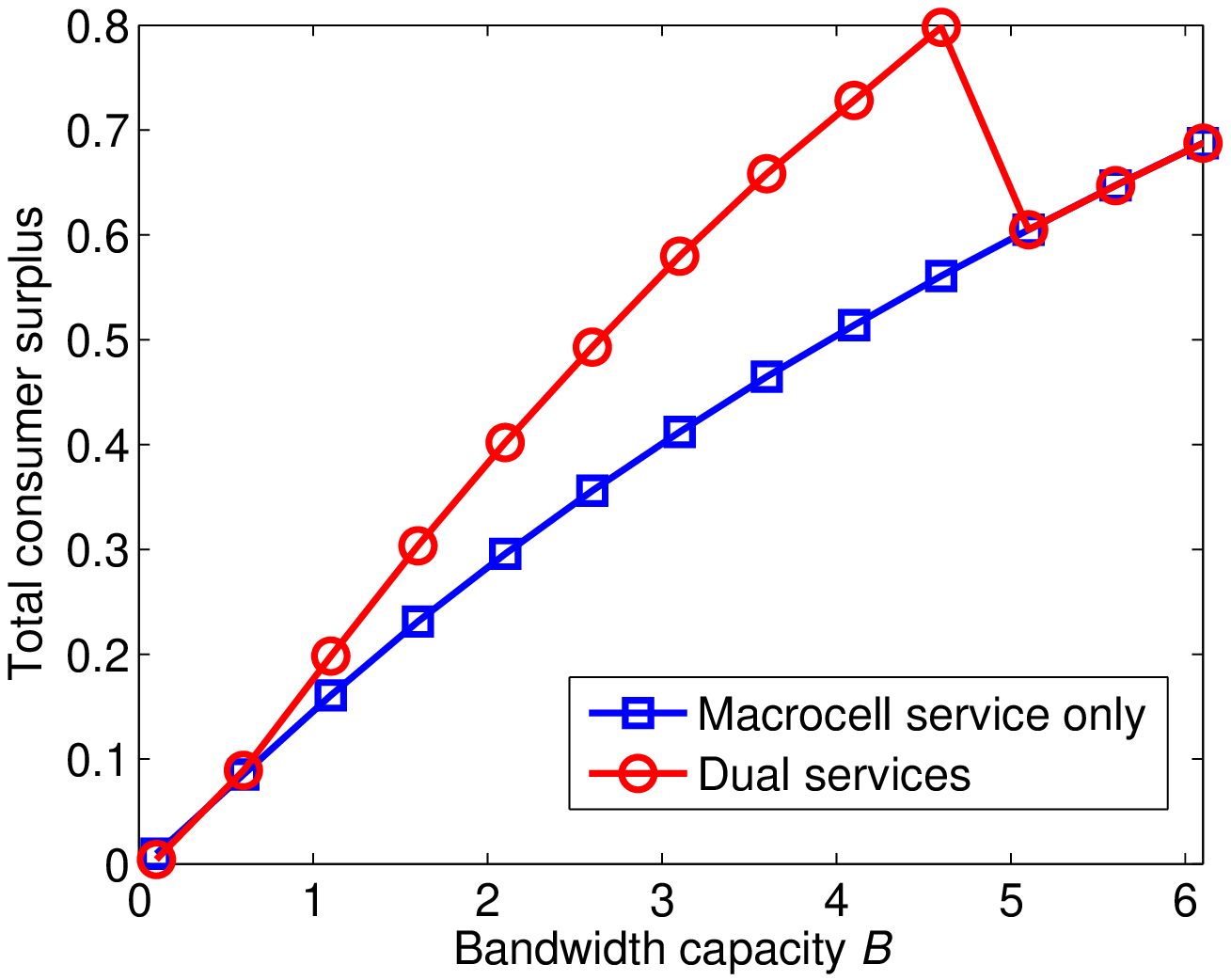}
      \caption{Comparison of total consumer surplus between dual services
and macrocell service only benchmark as functions of $B$.}
\label{fig:users_aggregate_B}
   \end{minipage}
   \begin{minipage}[t]{0.49\linewidth}
      \centering
      \includegraphics[width=1\textwidth]{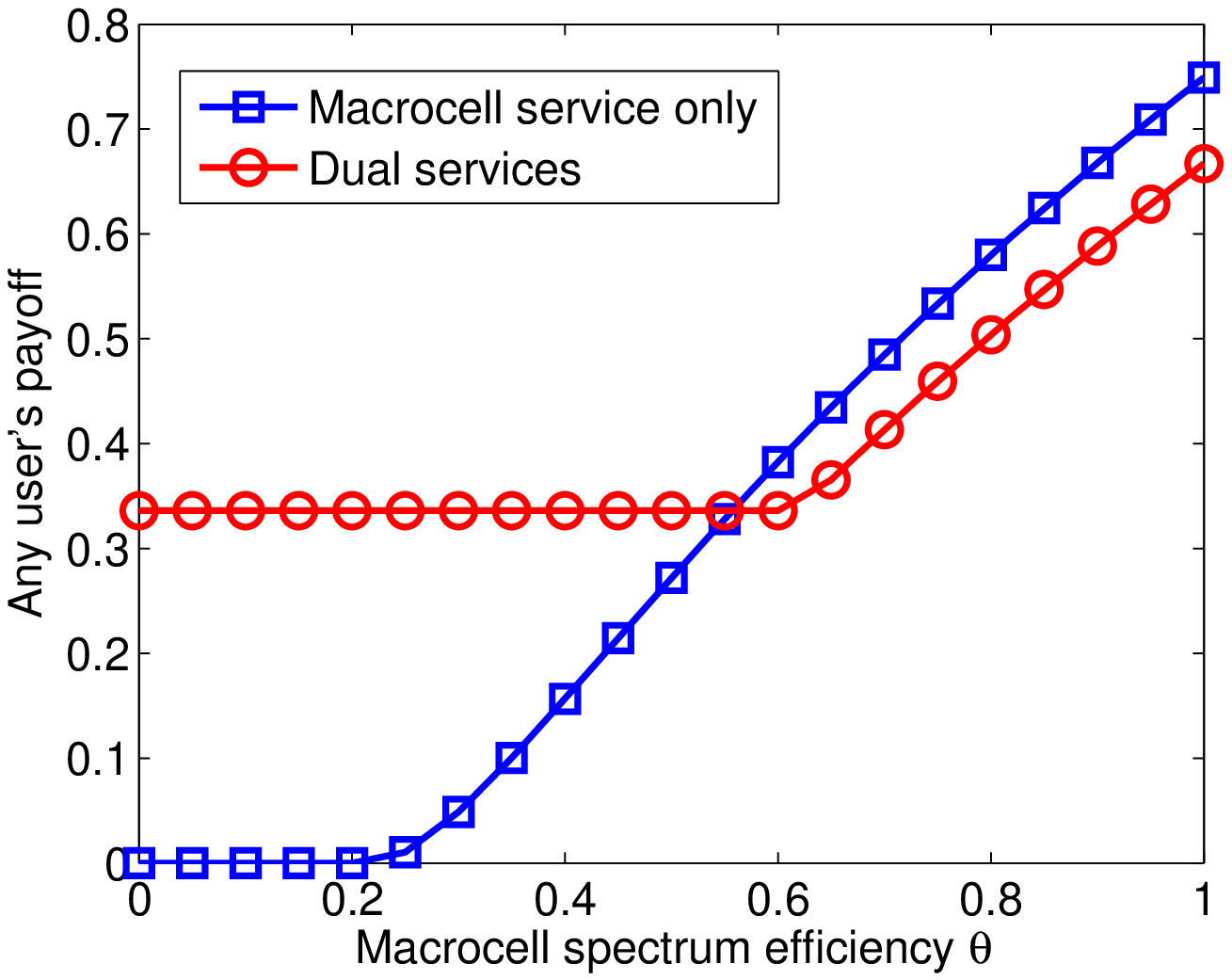}
      \caption{Comparison of a consumer's payoff between dual services and macrocell service only benchmark as functions of $\theta$. Here we fix $B=2.1$.}
\label{fig:UsersPayoff_theta}
   \end{minipage}
\end{figure*}


Figure~\ref{fig:users_aggregate_B} shows that the total consumer
surplus increases with the deployment of femtocell service. This is
mainly because users with low spectrum efficiency (small values
of $\theta$) will be able to obtain high {quality} service by using
femtocell and have better payoffs.
%

{ Nevertheless, some consumers could actually be worse off with dual
services. Figure~\ref{fig:UsersPayoff_theta} shows users' surplus
(payoff) with respect to their macrocell spectrum efficiency $\theta$.
Here we let the total capacity $B=2.1$. From
Figure~\ref{fig:users_aggregate_B} we know that, in this case, femtocell
deployment increases the total consumer surplus. However, Figure~\ref{fig:UsersPayoff_theta} shows that, for
individual users, a user
with small $\theta$ obtains great payoff enhancement with femtocell
deployment, whereas the payoff for a user with large $\theta$
actually becomes worse off. The former is due to the benefit of
service availability and quality improvement, while the latter is due to higher macrocell price. }

Based on Figs.~\ref{fig:unfair_distributed_profit_B} to \ref{fig:UsersPayoff_theta}, we have the following result.

\begin{observation}
After introducing femtocell service, the macrocell operator's profit increases since more users can be served now. Similarly, the total consumer surplus increases though some users' payoffs decrease. Overall, the total social welfare increases.
\end{observation}
}

\section{Extension I: With Femtocell Operational Cost}\label{sec:cost}

In Section~\ref{sec:UnfairFemto}, we consider a model where
femtocell service does not incur any additional cost comparing with
the macrocell service.  As shown in Figure~\ref{fig:coverage_femto},
femtocell users' traffic will first go through users' home wireline
broadband connections before reaching the control center of the
cellular network. The broadband connection is owned by an Internet
Service Provider (ISP). When the femtocell operator and the ISP
belong to the same entity (\eg both belonging to AT\&T) or the ISP
is sharing-friendly (NetShare 2002), there is no additional
cost for broadband access. Otherwise there is usually an
access charge. In this section, we study the more general case where
the ISP charges the femtocell operator fees for using the wireline
Internet connection.
We are interested in understanding how this
operational cost affects the provision of femtocell service.

For simplicity, we assume that the total operational cost is
linearly proportional to femtocell bandwidth with a coefficient
$C\in(0,1)$. {This is motivated by the fact that femtocell traffics will use ISP's broadband resource, and many ISPs have adopted usage-based pricing (Deleon 2011).} {If $C\geq 1$, we can show that the femtocell
operator will charge a femtocell price {$p_{F}>C\geq 1$}, and no
user will choose the femtocell service (see (\ref{eq:femtodemand})).
Thus we will focus on the case where the linear coefficient
$C\in(0,1)$.} The three-stage decision process is similar to that shwon in
Figure~\ref{fig:threestage_new}. The analysis of Stage III is the same
as in Section~\ref{subsec:StageIII}, and we hence focus on Stage II.

\subsection{Stage II: Femtocell Operator's Spectrum Purchase and Pricing Decisions}
In Stage II, the femtocell operator determines $B_R$ and $p_F$ to
maximize its profit. We still use $B_R^\ast(p_M,B_F)$ and
$p_F^\ast(p_M,B_F)$ to denote the equilibrium decisions of the
femtocell operator.

By following a similar analysis as in
Lemma~\ref{lemma:no_switchers}, we obtain the following result.

\begin{lemma}\label{lemma:no_switchers_cost}
At the equilibrium, the macrocell operator will satisfy all
preferred demands from users with $\theta\in[\theta_{th}^{pr},1]$.
\end{lemma}

The proof of Lemma~\ref{lemma:no_switchers_cost} is very similar to
Lemma~\ref{lemma:no_switchers} and is omitted. Based on
Lemma~\ref{lemma:no_switchers_cost}, the femtocell operator will
serve users with $\theta\in[0,{p_M}/{p_F})$, and its profit is
\begin{equation}\label{eq:femtocellprofitwithcost}
\pi^{Femto}(p_F,B_R)=(p_F-C)\min\left(B_R,\int_0^{\frac{p_M}{p_F}}\left(\frac{1}{p_F}-1\right)d\theta\right)-p_MB_R.
\end{equation}

we can explicitly write the femtocell operator's profit-maximization
problem as
\begin{align}\label{eq:opt_Distributed_Femto_cost}
&\max_{p_F\geq 0,B_R\geq 0}\ \ \pi^{Femto}(p_F,B_R)\nonumber\\
&\text{subject to} \ \ \
B_R\leq B_F,\nonumber\\
&\ \ \ \ \ \ \ \ \ \ \ \ \ \ p_M+C\leq p_F\leq 1,
\end{align}
where the second constraint shows that the femtocell price $p_F$
should at least cover the total cost ($p_M+C$) for the femtocell
operator. By solving Problem~(\ref{eq:opt_Distributed_Femto_cost}),
we obtain the following result.

\begin{lemma}\label{lemma:femto_cost}
In Stage II, the femtocell operator's equilibrium femtocell price is
\begin{equation}\label{eq:pF_Femto_cost}
p_{F}^{\ast}(p_M,B_F)=\max\left(\frac{2}{1+\frac{1}{p_M+C}},\frac{-p_M+\sqrt{(p_M)^2+4B_Fp_M}}{2B_F}\right),
\end{equation}
and its equilibrium femtocell bandwidth purchase is
\begin{equation}\label{eq:BR_Femto_cost}
B_{R}^{\ast}(p_M,B_F)=\min\left(p_M\frac{\frac{1}{(p_M+C)^2}-1}{4},B_F\right),
\end{equation}
which equals users' total preferred demand in femtocell service.
Then, users' preferred partition threshold equals users' finalized
partition threshold (\ie $\theta_{th}=\theta_{th}^{pr}$).
\end{lemma}

The proof of Lemma~\ref{lemma:femto_cost} is given in
Appendix~2.

\subsection{Stage I: Macrocell Operator's Spectrum Allocations and Pricing Decisions}

Now let us study Stage I, where the macrocell operator determines
$p_M$, $B_F$, and $B_M$ to maximize its profit. Let us denote its
equilibrium decisions as $p_M^\ast$, $B_F^\ast$, and $B_M^\ast$.

Notice that Lemma~\ref{lemma:no_switchers_cost} shows that it is
optimal for the macrocell operator to serve all users with
$\theta\in\left[\frac{p_M}{p_F^\ast(p_M,B_F)},1\right]$ by the
macrocell service.  By using the fact that $B_M=B-B_F$, we can
eliminate variable $B_M$. The macrocell operator's profit is
\begin{equation}
\pi^{Macro}(p_{M},
B_{F})=p_MB_{R}^*(B_F,p_M)+p_M\int_{\frac{p_M}{p_{F}^*(B_F,p_M)}}^1\left(\frac{1}{p_M}-\frac{1}{\theta}\right)d\theta.
\end{equation}
The macrocell operator's profit-maximization problem is
\begin{align}\label{eq:opt_cost_Macro_distr}
&\max_{B_F,p_M} \ \ \ \ \ \ \pi^{Macro}(p_{M}, B_{F}),
\nonumber \\
&\text{subject\ to}\ \  0\leq B_F+\int_{\frac{p_M}{p_{F}^*(B_F,p_M)}}^1\left(\frac{1}{p_M}-\frac{1}{\theta}\right)d\theta\leq B,\nonumber\\
& \ \ \ \ \ \ \ \ \ \ \ \  \ \ 0<p_M\leq 1-C,
\end{align}
where $B_{R}^*(B_F,p_M)$ and $p_{F}^*(B_F,p_M)$ are given in
(\ref{eq:pF_Femto_cost}) and (\ref{eq:BR_Femto_cost}), respectively.
The second constraint shows that the total cost $C+p_M$ to femtocell
operator should be less than $1$. Otherwise, the femtocell price
$p_F$ needs to be larger than $C+p_{M}$ and thus larger than 1, and
no user will subscribe to the femtocell service.


\subsection{Numerical Results}\label{sec:costnumerical}

Problem~(\ref{eq:opt_cost_Macro_distr}) is not convex and is
difficult to solve in closed-form, but can be solved easily using
numerical methods. Similar to Section~\ref{sec:UnfairFemto}, we can
see that
dual services degenerate to the macrocell service only benchmark
when capacity is large.
Here we will focus on how cost $C$ will affect the division of two
capacity regimes and the performance when capacity is small.

{
\subsubsection{Impact of $C$ on Capacity Regime Boundary}

Figure~\ref{fig:Boundary_BoverC} illustrates how cost $C$ affects
the boundary between the low capacity and high capacity regimes.
Recall that the boundary is 4.77 when $C=0$ (\ie
Figures~\ref{fig:unfair_distributed_bands_B},
\ref{fig:unfair_distributed_prices_B}, and
\ref{fig:unfair_distributed_partition_B}). When $C$ increases, the
femtocell price $p_{F}^{\ast}$ increases and demand for femtocell
service decreases. This makes it less attractive to provide
femtocell service. On the other hand, the increase of price
$p_{F}^{\ast}$ also reduces the market competition, which  makes the
macrocell operator more willing to lease spectrum to the femtocell
operator. The interactions of these two factors determine the
boundary of the two capacity regimes. More specifically, with a
small cost $C\leq 0.12$, the decrease of femtocell demands dominates
and the boundary decreases. With a large cost $C>0.12$, the decrease
of competition dominates and the boundary increases. We will discuss
these two factors in more details at a later point.

\begin{figure*}
   \begin{minipage}[t]{0.48\linewidth}
      \centering
      \includegraphics[width=0.8\textwidth,angle=90]{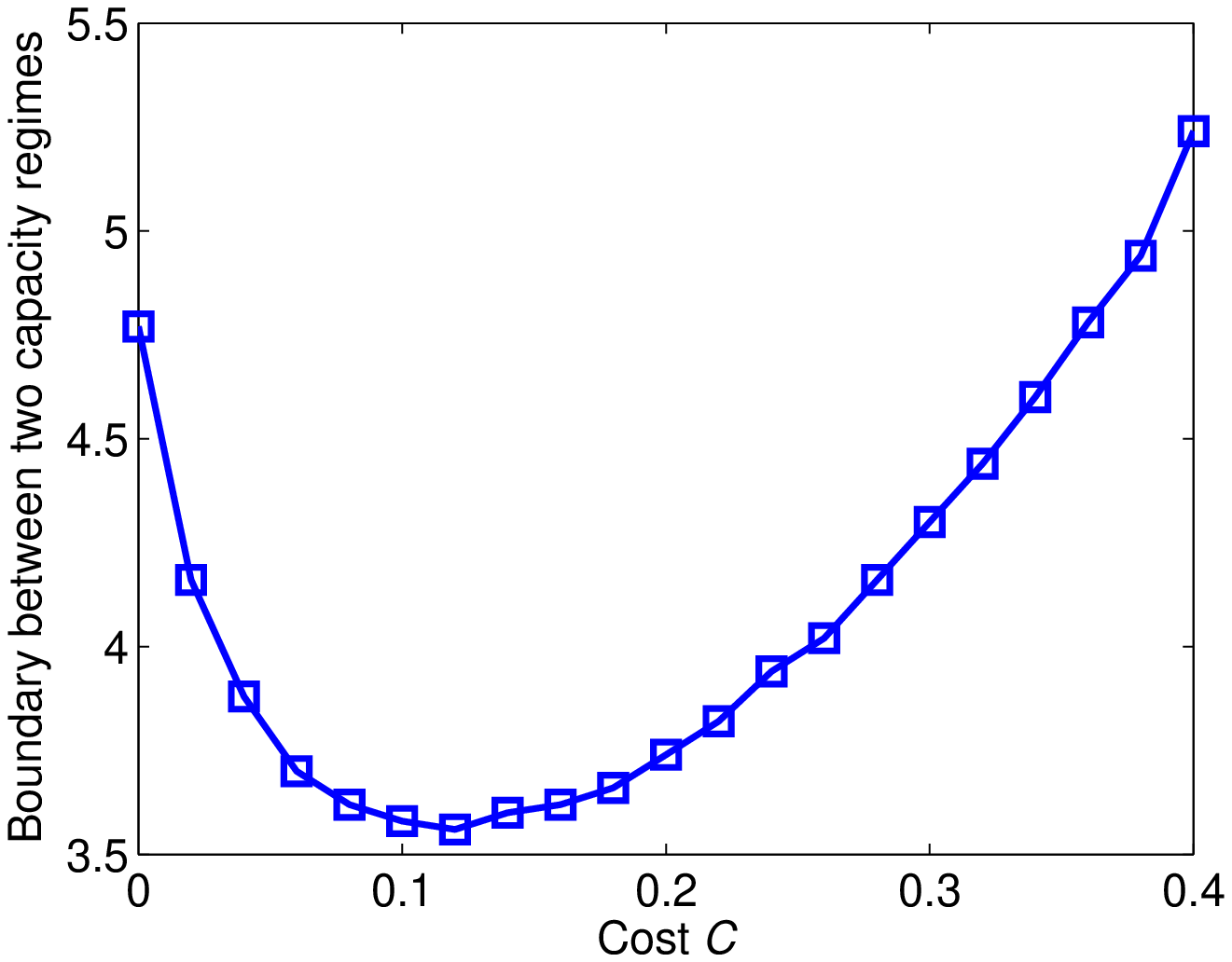}
\caption{The boundary between low and high capacity regimes change
with femtocell operational cost $C$.} \label{fig:Boundary_BoverC}
   \end{minipage}
   \begin{minipage}[t]{0.48\linewidth}
      \centering
      \includegraphics[width=0.8\textwidth,angle=90]{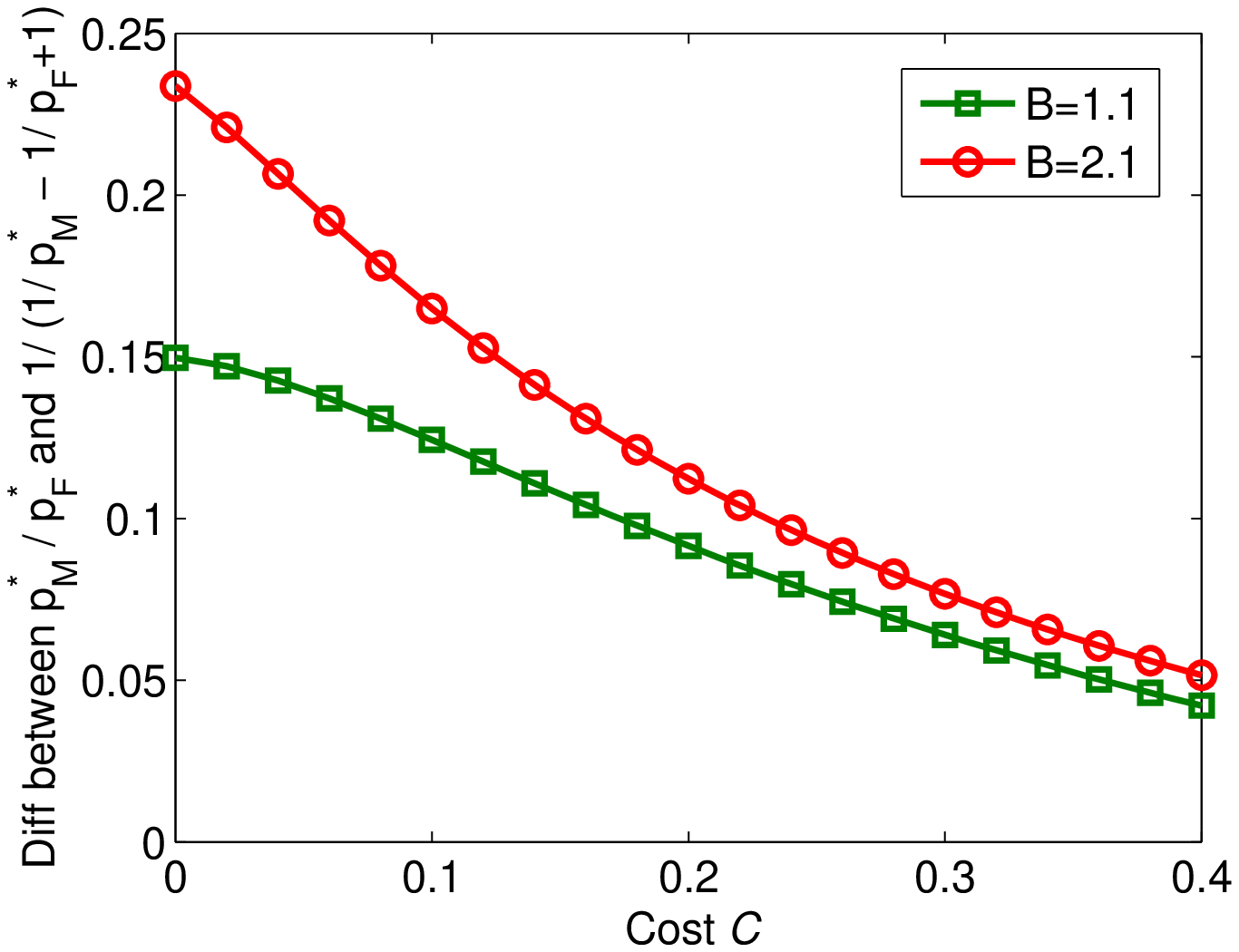}
\caption{The difference between users' partition
threshold $\theta_{th}=\frac{p_M^\ast}{p_F^\ast}$ and macrocell
operator's preferred threshold
$\tilde{\theta}_{th}=\frac{1}{\frac{1}{p_M^\ast}-\frac{1}{p_F^\ast}+1}$
as a function of $C$ and $B$}
\label{fig:partition_cost}
   \end{minipage}
\end{figure*}

Figure~\ref{fig:partition_cost} explicitly illustrates that a larger $C$ decreases
the gap between users' finalized partition threshold
$\theta_{th}=\frac{p_M^\ast}{p_F^\ast}$ and the threshold
$\tilde{\theta}_{th}=\frac{1}{\frac{1}{p_M^\ast}-\frac{1}{p_F^\ast}+1}$
that the macrocell operator prefers, and thus makes the service
competition less fierce. This gives more incentive to the macrocell
operator to lease spectrum to the femtocell operator, which is the
dominant factor that increases the boundary between two capacity
regions when $C$ increases (as shown in
Figure~\ref{fig:Boundary_BoverC}).

\begin{observation}
As cost $C$ increases in femtocells, the femtocell operator has less incentive to provide femtocell service. However, the macrocell operator may benefit from the increase of $C$ in terms of its profit since the service competition from femtocell operator become less intense.
\end{observation}

When cost $C$ increases but is still small, we can show that femtocell price increases to compensate cost, and the macrocell operator will face the decrease of femtocell demands. In this case, less femtocell band is needed and the macrocell operator's profit decreases in $C$ (see Fig.~\ref{fig:profits_distributed_cost}). However, when cost $C$ is large, competition between dual services mitigates and the macrocell operator allow the existence of femtocell service even for a large $B$. We can further observe from Fig.~\ref{fig:Boundary_BoverC} that the macrocell operator still wants to lease bandwidth to femtocell operator even when $B=5.2$ under $C=0.4$, while no femtocell service is provided under $C=0$ in this case. Thus the macrocell operator benefits from the high cost for large $B$ in terms of its profit.


%

\begin{figure*}
   \begin{minipage}[t]{0.49\linewidth}
      \centering
      \includegraphics[width=1\textwidth]{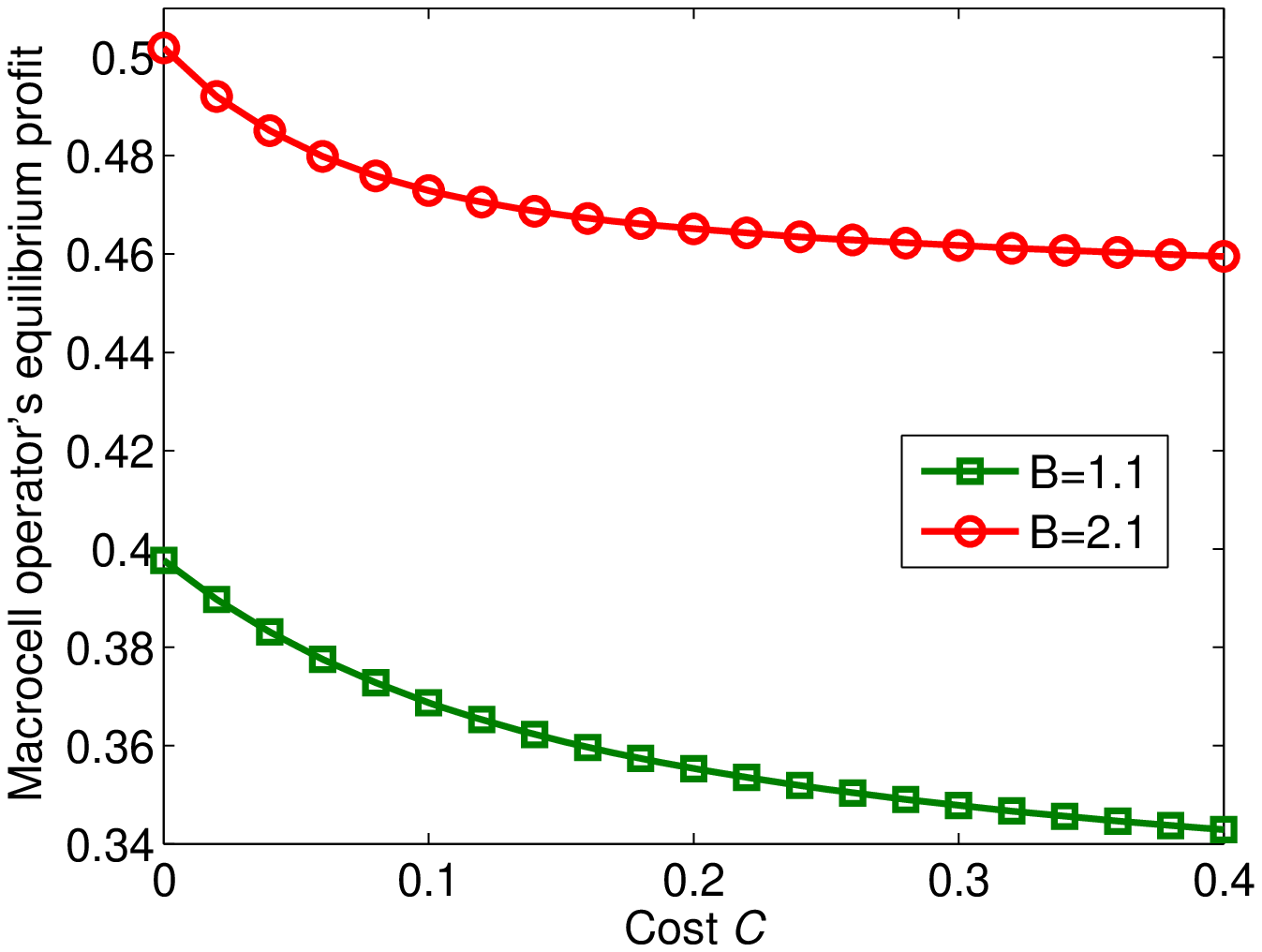}
\caption{Macrocell operator's equilibrium profit as a function of
cost $C$ and capacity $B$} \label{fig:profits_distributed_cost}
   \end{minipage}
   \begin{minipage}[t]{0.49\linewidth}
      \centering
      \includegraphics[width=1\textwidth]{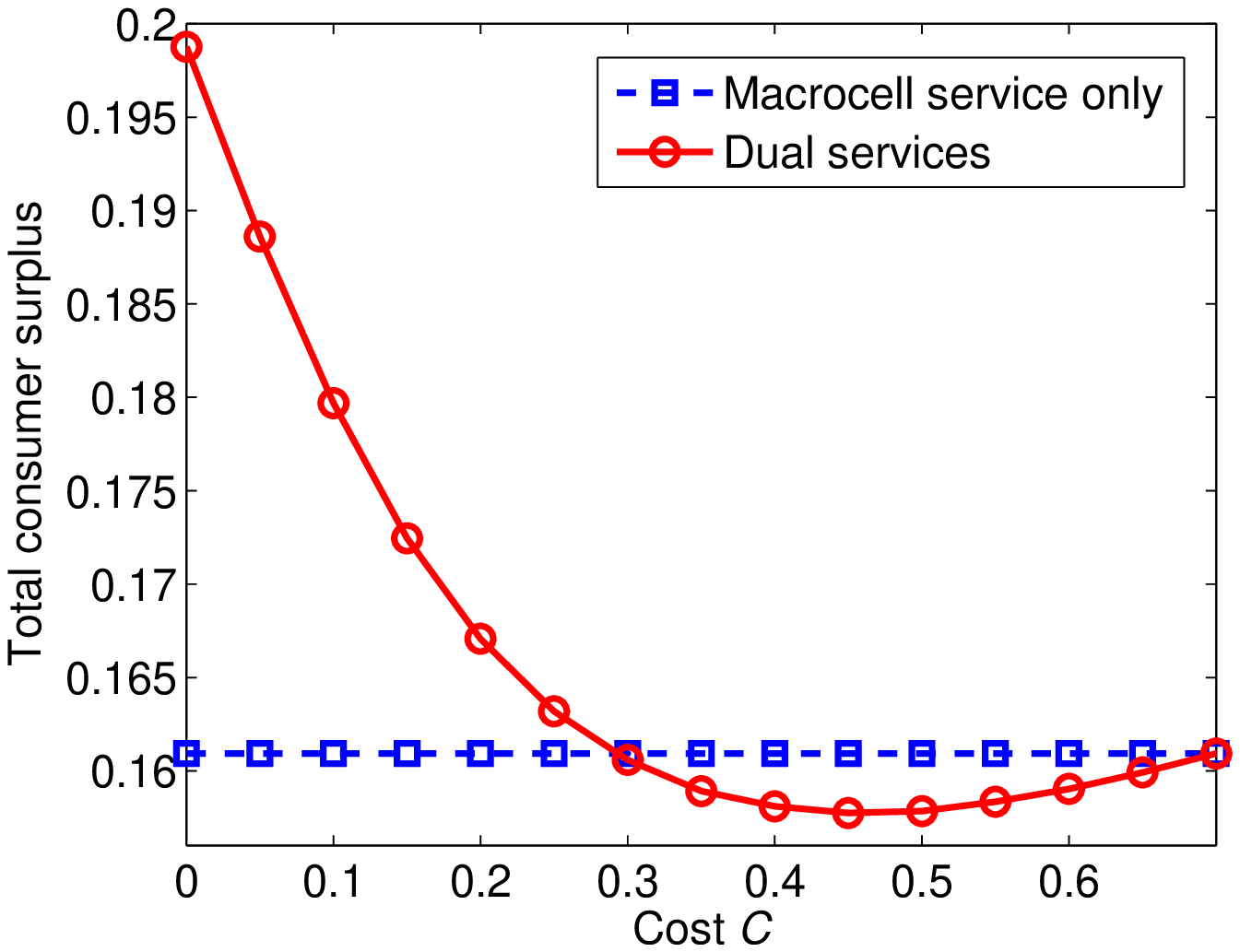}
\caption{Comparison of total consumer surplus between dual services
and macrocell service only benchmark as functions of $C$. Here
we fix $B=1.1$.} \label{fig:UsersPayoff_cost}
   \end{minipage}
\end{figure*}

}

\subsubsection{Impact of $C$ on Consumer Surplus and Social Welfare}
\begin{figure*}
   \begin{minipage}[t]{0.49\linewidth}
      \centering
      \includegraphics[width=0.8\textwidth]{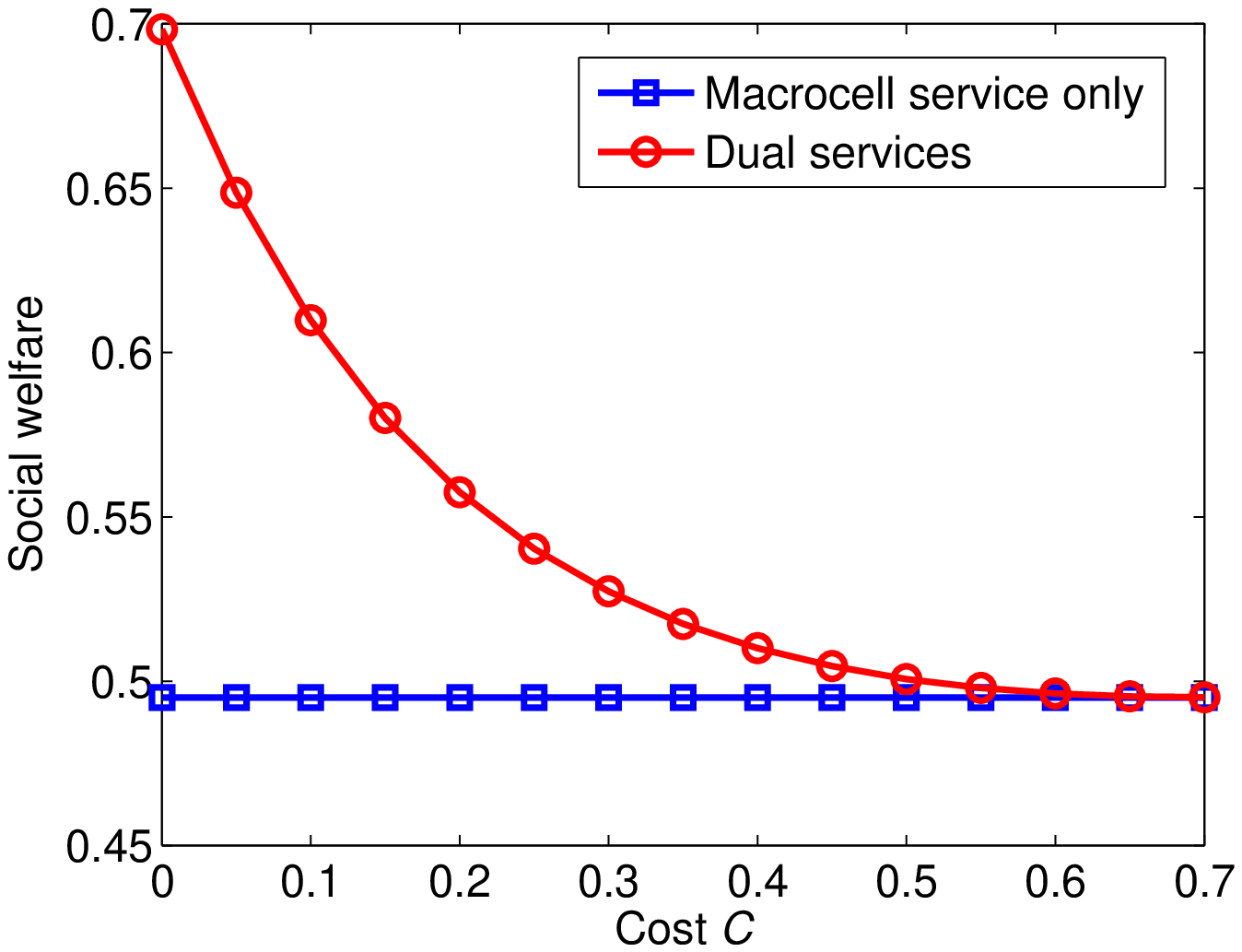}
\caption{Comparison of social welfare between dual services and
macrocell service only as functions of cost $C$. Here the total
capacity is fixed at $B=1.1$} \label{fig:Socialwelfare}
   \end{minipage}
   \begin{minipage}[t]{0.49\linewidth}
      \centering
      \includegraphics[width=1.05\textwidth]{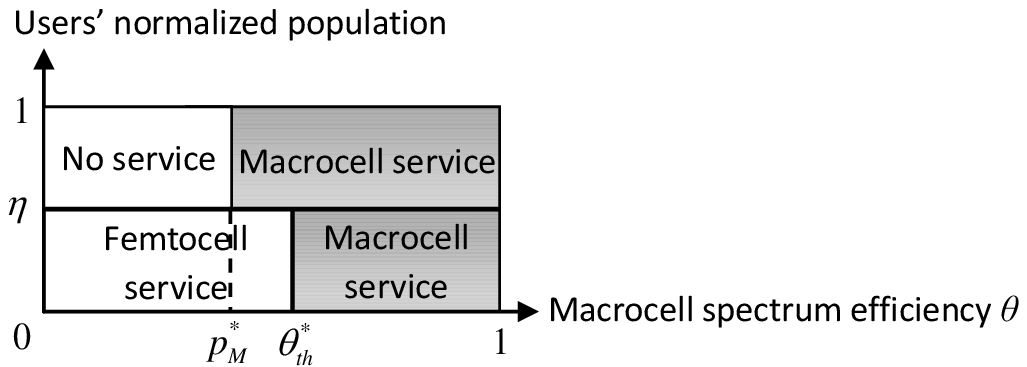}
\caption{Users' possible service partitions over space and macrocell
spectrum efficiency $\theta$} \label{fig:Coverage}
   \end{minipage}
\end{figure*}

In Figures~\ref{fig:UsersPayoff_cost} and \ref{fig:Socialwelfare},
we investigate how the cost $C$ affects total consumer surplus and
social welfare. We focus on the low capacity regime only.

Figure~\ref{fig:UsersPayoff_cost} shows that the total consumer
surplus is larger with dual services when $C<0.3$, but is smaller
with dual services when $C>0.3$. In the latter case, femtocell users
experience only small QoS improvements due to the high cost
$p_{F}^{\ast}$, and macrocell users experience a $p_M^\ast$ larger
than $p_M^{bench}$. {Note that macrocell price increases since all users can be served.} As a result, the total consumer surplus
decreases with dual services.

Figure~\ref{fig:Socialwelfare} shows that social welfare is always
larger with dual services for all possible values of $C$. Together
with Figure~\ref{fig:UsersPayoff_cost}, this shows that the macrocell
operator obtains a larger profit by sacrificing the consumer surplus
when $C>0.3$.

{
\begin{observation}
After introducing femtocell service, total consumer surplus increases only when the cost $C$ is small. The social surplus always increases.
\end{observation}
}

\section{Extension II: With Limited Femtocell Coverage}\label{sec:coverage}

In Section~\ref{sec:UnfairFemto}, we assume that femtocell service
has the same ubiquitous  coverage as the macrocell service. In this
section, we look at the general case where the femtocell service
only covers $\eta\in(0,1)$ portion of the user population, as
illustrated in Figure~\ref{fig:coverage_femto}. Then $1-\eta$ portion of users can only access the
macrocell service. Figure~\ref{fig:Coverage} illustrates the users'
possible service partitions over space and macrocell spectrum
efficiency $\theta$. We call the $\eta$ fraction users
\emph{overlapped users}, and the rest $1-\eta$
\emph{non-overlapped users}.
We are interested in understanding how the limited coverage affects
the provision of femtocell service.

The three-stage decision process is similar to that depicted in
Figure~\ref{fig:threestage_new}. The analysis of Stage III is the same
as Section~\ref{subsec:StageIII}. Next we focus on Stage II.

Following a similar analysis as in Lemma~\ref{lemma:no_switchers},
we can also conclude that overlapped users with
$\theta\in[\theta_{th}^{pr},1]$ will be served by macrocell service,
and the other overlapped users will be served by femtocell service.
That is, $\theta_{th}=\theta_{th}^{pr}$. Then we can similarly
derive the following result.



\begin{lemma}\label{lemma:Femto}
In Stage II, the femtocell operator's equilibrium femtocell price is
\begin{equation}\label{eq:p_F_coverage}
p_{F}^\ast(p_M,B_F)=\max\left(\frac{2}{\frac{1}{p_M}+1},\frac{-p_M\eta+\sqrt{(p_M\eta)^2+4p_MB_F\eta}}{2B_F}\right),
\end{equation}
and its leased bandwidth from macrocell operator is
\begin{equation}\label{eq:B_R_coverage}
B_{R}^\ast(p_M,B_F)=\min\left(\eta
p_M\left(\frac{\frac{1}{(p^M)^2}-1}{4}\right),B_F\right),
\end{equation}
which equals overlapped users' total preferred demand in femtocell
service.
\end{lemma}

%

\subsection{Macrocell Operator's Spectrum Allocations and Pricing in Stage I}
Now we are ready to study Stage I, where the macrocell operator's
profit-maximization problem is
%
%
\begin{align}\label{eq:opt_Macro_coverage}
&\max_{p_M,B_F}\pi^{Macro}(p_M,B_F)=p_MB_R^\ast(p_M,B_F)+p_M\int_{\frac{p_M}{p_F^\ast(p_M,B_F)}}\left(\frac{1}{p_M}-\frac{1}{\theta}\right)d\theta,\nonumber\\
&\text{subject\ to,}\ \ \ \ \ \ \ \ \ \
0<B_F+\int_{\frac{p_M}{p_F^\ast(p_M,B_F)}}\left(\frac{1}{p_M}-\frac{1}{\theta}\right)d\theta\leq
B,
\end{align}
where $p_{F}^\ast(p_M,B_F)$ and $B_{R}^\ast(p_M,B_F)$ are
respectively given in (\ref{eq:p_F_coverage}) and
(\ref{eq:B_R_coverage}).

\subsection{Numerical Results}
Problem~(\ref{eq:opt_Macro_coverage}) is not convex and is difficult
to solve in closed-form, but can be solved easily using numerical
methods. As in Sections~\ref{sec:unfairfemtollnum} and
\ref{sec:costnumerical}, we can again clearly observe different
behaviors in two capacity regimes: dual services degenerate to the
macrocell service only benchmark in the high capacity regime.
Unlike Section~\ref{sec:costnumerical},  the femtocell
coverage $\eta$ does not affect the boundary of the two capacity
regions (\ie always at $B=4.77$). {The two effects (QoS improvement and competition brought by femtocell service) coexist in $\eta$ coverage.}


%

\begin{figure*}
   \begin{minipage}[t]{0.49\linewidth}
      \centering
      \includegraphics[width=0.9\textwidth]{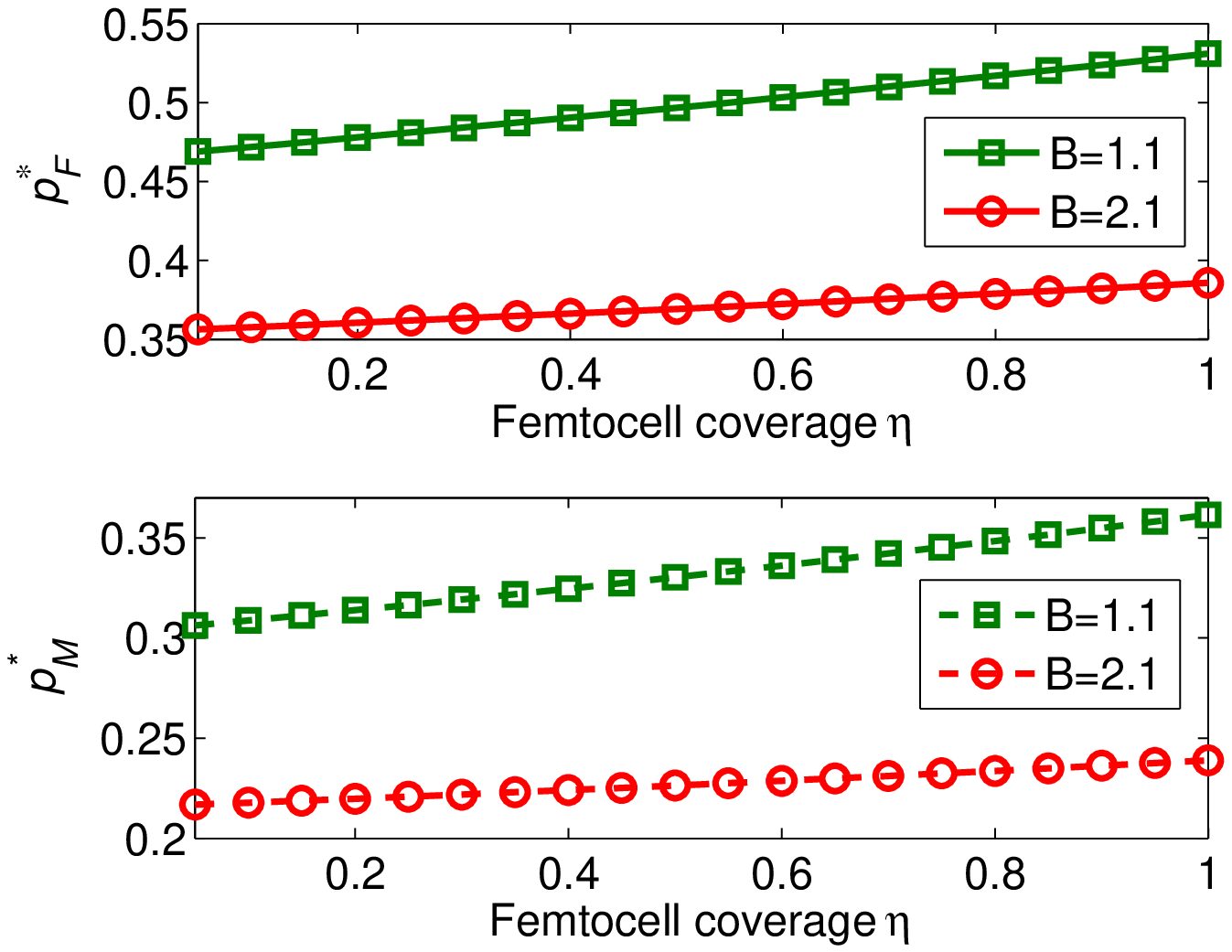}
\caption{Equilibrium prices $p_{F}^{\ast}$ and $p_{M}^{\ast}$ as functions of $\eta$ and
$B$} \label{fig:prices_distributed_coverage}
   \end{minipage}
   \begin{minipage}[t]{0.49\linewidth}
       \centering
       \includegraphics[width=0.9\textwidth]{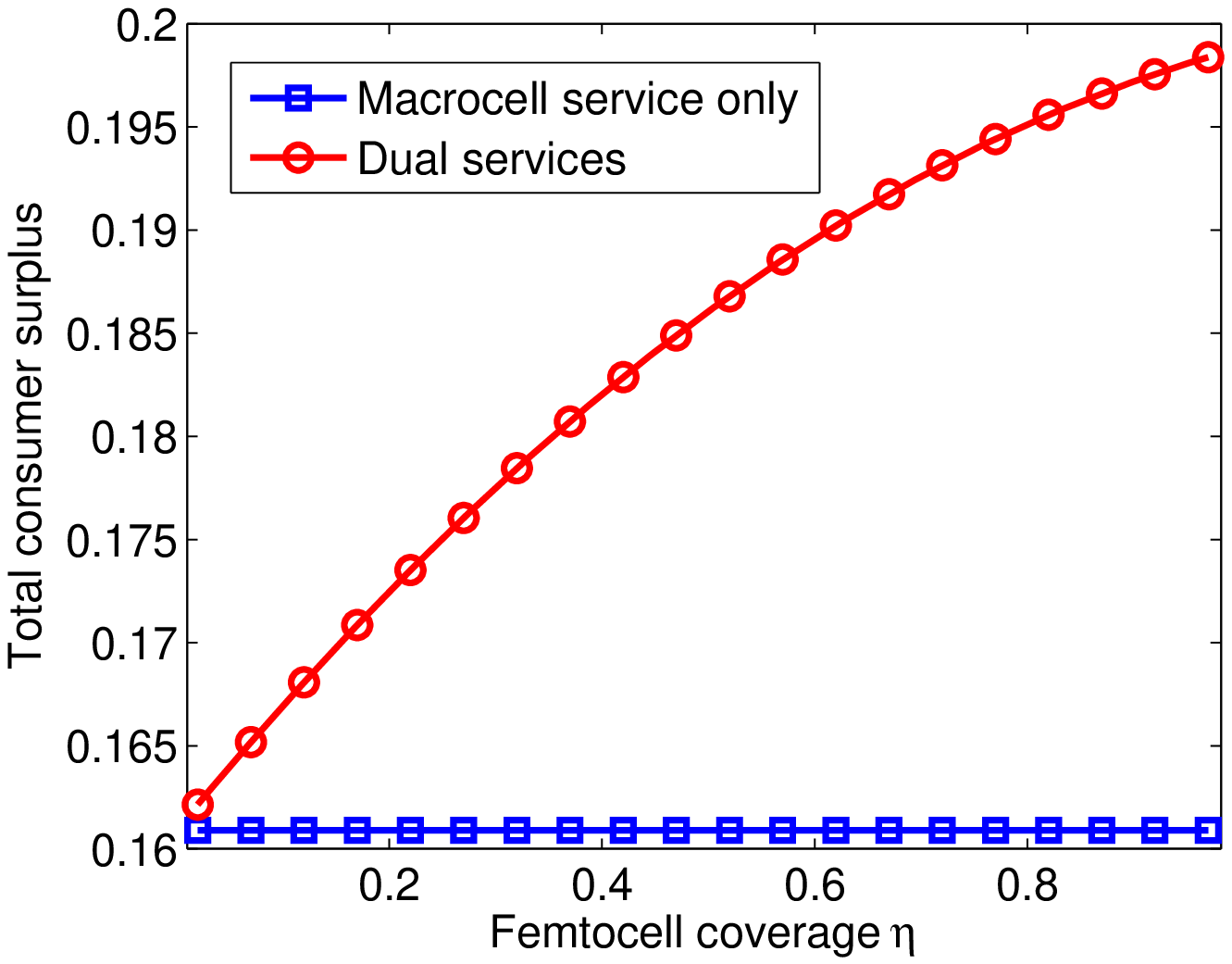}
\caption{Comparison of total consumer surplus between dual services
and macrocell benchmark under $B=1.1$.}
\label{fig:ConsumerPurlus_Eta}
       \end{minipage}
\end{figure*}

{We can show that as $\eta$ increases, it is more attractive to provide femtocell service and the equilibrium femtocell (macrocell) band $B_F^*$ ($B_M^*$) increases (decreases). Yet both prices $p_F^*$ and $p_M^*$ increase in $\eta$ (see Fig.~\ref{fig:prices_distributed_coverage}). Intuitively, as $\eta$ increases, more users are served with a larger total demand in femtocell service, which leads to a larger $p_F^*$. The overall wireless service (macrocell plus femtocell) become more efficient and the total user demand (of both services) will increase. Thus we can observe a larger $p_M^*$. Since the macrocell operator can sell total capacity in higher dual prices, its profit increases in $\eta$. }
%



{Figure~\ref{fig:ConsumerPurlus_Eta} further shows that the total consumer
surplus is larger with dual services than macrocell service only
benchmark. This result is similar to Fig.~\ref{fig:users_aggregate_B} in
Section~\ref{sec:unfairfemtollnum}.
%

\begin{observation}
As femtocell coverage expands, the overall wireless service (macrocell plus femtocell) becomes more efficient. Both macrocell operator's profit and total consumer surplus (as well as social welfare) increase in $\eta$.
\end{observation}}

\section{Conclusion}\label{sec:conclusion}
{This paper studies the economic incentives for a macrocell
operator to deploy new femtocell service in addition to its existing
macrocell service. The femtocell service is provided by another
party, the femtocell operator, who needs to lease the macrocell
operator's capacity. We model the interactions among macrocell
operator, femtocell operator, and users as a three-stage dynamic
game, and derive the equilibrium capacity allocation and pricing decisions. Our analysis shows that the macrocell operator has an
incentive to enable both macrocell and femtocell services when its
total bandwidth is small, as femtocell service enhances user
coverage and improves profits for both macrocell and femtocell
operators. Notice that not all users will experience a payoff
increase by the introduction of femtocell service in this case.
However, when the total bandwidth is large, femtocell service becomes
a severe competitor to macrocell service, and the macrocell
operator thus has less incentive to lease its bandwidth to the femtocell
operator. In this case, only macrocell service is provided to users.

Also, we further study the impact of operational cost of femtocell
service. On one hand, we show that the increase of operational cost
of femtocell service makes both operators' profits decrease. On the
other hand, we show that the operational cost can mitigate femtocell
operator's competition with macrocell operator. Finally, we
investigate the impact of limited femtocell coverage where only some users have access to femtocell service.}

There are several directions to extend the results in this
paper.
\begin{itemize}
\item We can further consider the ``shared carriers'' scheme besides ``separate carriers'', where
femtocell service and macrocell service share part of or the whole
spectrum. We need to optimize the pricing and spectrum allocation
decisions by trading off the increased spectrum efficiency and
mutual interferences between macrocell and femtocell services.
\item We can also
consider the frequency spectrum reuse, where multiple femtocells can
reuse the same spectrum if they do not overlap with each other in
terms of coverage. In this case, femtocell service will become more
attractive to the femtocell operator, as a single frequency band may
support more users. However, frequency reuse might make the
interference management complicated in areas where femtocells are
densely deployed.
\item {We may also consider a more practical users' models in dual services, by incorporating their heterogeneous levels of willingness to pay or sensitivity to achieved data rates.}
\item {Moreover, we can extend our current monopoly case to oligopoly case, where multiple macrocell (femtocell) operators compete with each other. Intuitively, we can envision no provision of femtocell service only when all macrocell operators have adequate bandwidth. But the macrocell and femtocell prices will go down due to operators' competitions.}
\end{itemize}

{
\section*{Appendix}

\textbf{Appendix 1: Proof of Lemma~\ref{lemma:femto}}

We first
notice that the first term in the $\min$ operation of
$\pi^{Femto}(p_F,B_R)$ in (\ref{eq:femtocellProfit}) is increasing
in both $p_F$ and $B_R$, while the second term is decreasing in both
$p_F$ and $B_R$. Hence, the equilibrium $p_F^*(p_M,B_F)$ and
$B_R^*(p_M,B_F)$ should make these two terms equal (\ie femtocell
operator's capacity equals users' total preferred femtocell demand).
Then we can write $B_R^*(p_M,B_F)$ as a function of $p_F$, \ie
\begin{equation}\label{eq:BR_pF}
B_{R}^*(p_F)=\frac{(1-p_F)p_M}{(p_F)^2},
\end{equation}
which should be no larger than $B_F$ by a proper choice of $p^F$.

Then the femtocell operator's profit-maximization problem can be
simplified from (\ref{eq:opt_Distributed_Femto}) to
\begin{align}\label{eq:opt_Distributed_Femto2}
&\max_{p_F}\pi^{Femto}(p_F)=\pi^{Femto} (p_F,B_{R}^*(p_F))=\frac{p_M(1-p_F)(p_F-p_M)}{(p_F)^2}\nonumber\\
&\text{subject  to}\ \ \ \ \ \ \ \ \ \ \ \
\frac{(1-p_F)p_M}{(p_F)^2}\leq B_F.
\end{align}
Let us check the first derivative of $\pi^{Femto}(p_F)$ over $p_F$,
\ie
$$\frac{d\pi^{Femto}(p_F)}{dp_F}=p_M\frac{2p_M-(1+p_M)p_F}{(p_F)^3},$$
which is positive when $p_F<{2p_M}/{(1+p_M)}$ and is negative
otherwise. Thus $\pi^{Femto}(p_F)$ reaches its maximum when
$p_F={2p_M}/{(1+p_M)}$. Also, $p_F$ needs to satisfy the constraint
in (\ref{eq:opt_Distributed_Femto2}). Then the equilibrium femtocell
price is
$$p_{F}^{\ast}(p_M,B_F)=\max\left(\frac{2p_M}{1+p_M},\frac{-p_M+\sqrt{(p_M)^2+4B_Fp_M}}{2B_F}\right).$$
By substituting this into (\ref{eq:BR_pF}), we obtain the
equilibrium femtocell bandwidth purchase as
$$B_{R}^{\ast}(p_M,B_F)=\min\left(\frac{1-(p_M)^2}{4p_M},B_F\right).$$

\textbf{Appendix 2: Proof of
Lemma~\ref{lemma:femto_cost}}

We can
rewrite the femtocell operator's profit in
(\ref{eq:opt_Distributed_Femto_cost}) as
$$\pi^{Femto}(B_R,p_F)=\min\left(S_F(B_R,p_F),Q_F(B_R,p_F)\right),$$
where
$$S_F(B_R,p_F)=B_R(p_F-p_M-C)$$
and
$$Q_F(B_R,p_F)={\frac{p_M}{p_F}}\left(\frac{1}{p_F}-1\right)(p_F-C)-p_MB_R.$$
It is clear that $S_F(B_R,p_F)$ is increasing in $B_R$ and $p_F$,
and $Q_F(B_R,p_F)$ is decreasing in $B_R$.

Let us consider the following two cases at the equilibrium:
\begin{itemize}
\item If $S_F(B_R,p_F)>Q_F(B_R,p_F)$, the femtocell operator's
profit equals $Q_F(B_R,p_F)$. Then the femtocell operator will
decrease $B^R$, which will decrease $S_F(B_R,p_F)$ and increase
$S_F(B_R,p_F)$ until $S_F(B_R,p_F)=Q_F(B_R,p_F)$. The profit will be
improved. Thus  $S_F(B_R,p_F)>Q_F(B_R,p_F)$ cannot be true at the
equilibrium.

\item Similarly, we can show that it is not possible to have $S_F(B_R,p_F)<Q_F(B_R,p_F)$ at the equilibrium.
\end{itemize}
As a result, $S_F(B_R,p_F)=Q_F(B_R,p_F)$ at the equilibrium, which
leads to
\begin{equation}\label{eq:BR_pF}
B_{R}^*(p_F)=\frac{p_M}{p_F}\left(\frac{1}{p_F}-1\right).
\end{equation}
The choice of $p_F$ should satisfy that $B_{R}^*(p_F)<B_F$, \ie
\begin{equation}\label{eq:pFBF}
p_F\geq \frac{-p_M+\sqrt{(p_M)^2+4p_MB_F}}{2B_F}.
\end{equation}
Then the femtocell operator's profit-maximization problem in
(\ref{eq:opt_Distributed_Femto_cost}) can be simplified as
\begin{align}\label{proof:cost}
&\max_{p_F}\ \ \pi^{Femto}(p_F)=\frac{p_M}{p_F}\left(\frac{1}{p_F}-1\right)(p_F-p_M-C),\nonumber\\
&\text{subject\ to}\ \ \ p_F\geq
\frac{-p_M+\sqrt{(p_M)^2+4p_MB_F}}{2B_F}.
\end{align}

The first order derivative of $\pi^{Femto}(p_F)$ in
(\ref{proof:cost}) over $p_F$ is
$$\frac{d \pi^{Femto}(p_F)}{d p_F}=\frac{p_M}{(p_F)^2}\left(\left(C+p_M\right)\left(\frac{2}{p_F}-1\right)-1\right),$$
which is positive when $p_M<\frac{2}{1+\frac{1}{p_M+C}}$ and
negative otherwise. Thus $\pi^{Femto}(p_F)$ achieves its maximum at
$p_F=\frac{2}{1+\frac{1}{p_M+C}}$. Also, $p_F$ needs to satisfy the
constraint in Problem~(\ref{proof:cost}). Hence, we obtain the
equilibrium femtocell price $p_M^\ast(p_M,B_F)$ in
(\ref{eq:pF_Femto_cost}). By substituting (\ref{eq:pF_Femto_cost})
into (\ref{eq:BR_pF}), we can derive the equilibrium femtocell
bandwidth request $B_R^\ast(p_M,B_F)$ in (\ref{eq:BR_Femto_cost}).
}

%

\section*{References}

\begin{hangref}

\item
Altmann, J., K. Chu. 2001. How to charge for network services-flat-rate or usage-based? \emph{Computer Networks} \textbf{36(5-6)}~519--531.

\item
Atkinson, J. 2011. Vodafone and BT mobile mvno renew contract for five years.
http://www.\\mobiletoday.co.uk/News/11089/Vodafone\_and\_BT\_Mobile\_MVNO\_renew\_contract\_
for\_\\five\_years\_.aspx.

\item
AT\&T, Frequently Asked Questions. 2011. AT\&T 3g micocell. http://www.att.com/shop/\\wireless/
devices/3gmicrocell.jsp?fbid=2RqBuNua27C\#faqs.


\item
Berg Insight. 2009. {Femtocells and fixed-mobile convergence}.
{http://www.berginsight.com/\\ReportPDF/ProductSheet/bi-fmc-ps.pdf}.



\item
Chen, Y., J. Zhang, Q. Zhang. 2011. Utility-aware refunding framework for hybrid access femtocell network.
\emph{IEEE Transactions on Wireless Communications}.

\item
Chiang, W. K., D. Chhajed, J. D. Hess. 2003. Direct marketing,
indirect profits: a strategic analysis of dual-channel supply-chain
design. \emph{Management Science} \textbf{49(1)}~1--20.

\item
China Femtocell Symposium. 2011, Beijing. http://www.conference.cn/femto/2011/en/
Conference.asp?ArticleID=475.

\item
Claussen, H., L. Ho, L. Samuel. Financial analysis of a
pico-cellular home network deployment. \emph{IEEE ICC}.

\item
Courcoubetis, C., R. Weber. 2003. \emph{Pricing communication
networks- economics, technology and modeling}. New York: Wiley.

\item
Deleon, N. 2011. Usage-based billing hits Canada: say goodbye to Internet innovation. TechCrunch Hot Topics. http://techcrunch.com/2011/02/01/usage-based-billing-hits-canada-say-goodbye-to-internet-innovation/

\item
Dewenter, R., J. Haucap. 2006. Incentives to lease mobile virtual network operators (mvnos). presented at
the 34th Res. Conf. Commun., Inf. Internet Policy.

\item
Duan, L., J.~Huang, B. Shou. 2011. Economic incentives of femtocell service
provision. submitted to \emph{IEEE Transactions on Mobile Computing}.


\item
Dumrongsiri, A., M. Fan, A. Jain, K. Moinzadeh. 2008. A supply chain
model with direct and retail channels. \emph{European Journal of
Operational Research}~\textbf{187(3)} 691-718.

\item
Fitchard, K. 2009.
{http://connectedplanetonline.com/wireless/news/sprint-femtocells-whole\\sale-services-0610}.

\item
Goldstein, P. 2011. Verizon confirms it will ditch unlimited
smartphone data plans starting July 7. Fierce Wireless.
{http://www.fiercewireless.com/story/verizon-confirms-it-will-ditch-unlimited-smartphone-data-plans-starting-jul/2011-07-05}.

\item
Hong, W., Z. Tsai. 2010. On the femtocell-based MVNO model: a game
theoretic approach for optimal power setting. \emph{IEEE 71st
Vehicular Technology Conference (VTC)}.

\item
Huang, W., J. M. Swaminathan. 2009. Introduction of a second
channel: implications for pricing and profits. \emph{European
Journal of Operational Research} \textbf{194(1)}~258--279.

\item
Informa Telecoms \& Media. 2011. {Femtocell deployments more than
double in 12 months with strong growth byond the home}.
{http://www.informatm.com/itmgcontent/icoms/s/\\press-releases/20017850662.html}.
%

\item
LaVallee, A. 2009. AT\&T to New York and San Francisco: we are working on it. \emph{The Wall Street Journal}.


\item
McKnight, L. W., J. P. Bailey. 1998. \emph{Internet economics}. MIT
Press Cambridge, Massachusetts.

\item
Myerson, R. B. 2002. \emph{Game theory: analysis of conflict}.
Cambridge, MA: Harvard University Press.

\item
NetShare Speakeasy. 2002.
{http://www.prnewswire.com/news-releases/speakeasy-breaks-with-mainline-isps-embraces-wi-fi-and-promotes-shared-broadband-connections-76903267.html}


\item
Rubin, P. H.. 1978. The theory of the firm and the structure of the
franchise contract. \emph{The Journal of Law and Economics}
\textbf{21(1)}~223--233.

\item
Sandler, Kathy. 2009. House calls: `femtocells' promise to boost the cellphone signals inside your home. \emph{The
Wall Street Journal}.

\item
Sengupta, S., M., Chatterjee. 2009. An economic framework for
dynamic spectrum access and service pricing. {\it ACM/IEEE
Transactions on Networking\/} {\bf 17(4)} 1200--1213.

\item
Shetty, N., S. Parekh, J. Walrand. 2009. Economics of femtocells.
\emph{IEEE GLOBECOM}.


\item
Tanneur, E. P. L. 2003. Residential resale of wireless broadband via
wireless. Thesis at Ecole Polytechnique.

\item
Tsay, A. A., N. Agrawal. 2004. Channel conflict and coordination in
e-commerce age. \emph{Production and Operations Management}
\textbf{13(1)}~93--110.


\item
Wang, W., B. Li. 2005. Market-driven bandwidth allocation in selfish
overlay networks. \emph{IEEE INFOCOM}.

\item
WNN Wi-Fi Net News. 2008. Verizon wins big 700MHZ auction block.
{http://wifinetnews.com\\/archives/2008/03/verizon\_wins\_big\_700\_mhz\_auction\_block.html}.

%

\item
Yun, S., Y. Yi, D. Cho, J. Mo, Open or close: on the sharing of
femtocells. \emph{IEEE INFOCOM Mini-Conference}.

\end{hangref}

\end{document}